\documentclass[aps,prd,11pt,twoside,tightenlines,nofootinbib,showpacs,preprint,superscriptaddress]{revtex4-1}
\usepackage[colorlinks=true]{hyperref}
\usepackage{xcolor}
\hypersetup{colorlinks=true, citecolor=blue, urlcolor=blue, linkcolor=blue}
\usepackage{amsmath,amssymb}
\usepackage[final]{graphicx}
\usepackage{subcaption}
\usepackage{makecell}

\begin{document}

\title{Is two-poles' $\Lambda(1405)$ one state or two?}

\author{Zhong-Yu Wang}
\affiliation{School of Physics and Electronics, Hunan Key Laboratory of Nanophotonics and Devices, Central South University, Changsha 410083, China}

\author{Hiwa A. Ahmed}
\affiliation{School of Physics and Electronics, Hunan Key Laboratory of Nanophotonics and Devices, Central South University, Changsha 410083, China}
\affiliation{Physics Department, College of Science, University of Sulaimani, Kurdistan Region 46001, Iraq}

\author{C. W. Xiao}
\email{xiaochw@csu.edu.cn}
\affiliation{School of Physics and Electronics, Hunan Key Laboratory of Nanophotonics and Devices, Central South University, Changsha 410083, China}

\date{\today}
\begin{abstract}
To understand the nature of two poles for the $\Lambda(1405)$ state, we revisit the interactions of $\bar{K}N$ and $\pi\Sigma$ with their coupled channels, where two-poles structure is found in the second Riemann sheet. We also dynamically generate two poles in the single channel interaction of $\bar{K}N$ and $\pi\Sigma$, respectively. Moreover, we make a further study of two poles' properties by evaluating the couplings, the compositeness, the wave functions, and the radii for the interactions of four coupled channels, two coupled channels and the single channel. Our results show that the nature of two poles is unique. The higher-mass pole is a pure $\bar{K} N$ molecule, and the lower-mass one is a compositeness of mainly $\pi \Sigma$ with tiny component $\bar{K} N$. From our results, one can conclude that the $\Lambda(1405)$ state would be overlapped with two different states of the same quantum number. 
\end{abstract}
\maketitle

\section{Introduction}

Quantum chromodynamics (QCD) has become the fundamental theory for describing the strong interaction between quarks and gluons, and perturbative QCD has also achieved great success in explaining the strong interactions in the high energy region. However, in the low energy region where many resonances appear, the non-perturbative QCD should be applied to understand the properties of the hadronic states under the confinement. There are many non-perturbative QCD approaches proposed, seen more details in the reviews~\cite{Chen:2016qju,Hosaka:2016pey,Guo:2017jvc} and references therein. In recent years, due to the development of the high energy experiments, many exotic states were found, such as tetraquark-like or pentaquark-like states, which were a chance and challenge to understand the features of the non-perturbative QCD and caught much attentions both in theories and experiments, see the reviews~\cite{Chen:2016qju,Hosaka:2016pey,Guo:2017jvc,Chen:2016spr,Lebed:2016hpi,Esposito:2016noz,Ali:2017jda,Briceno:2017max,Olsen:2017bmm,Karliner:2017qhf,Yuan:2018inv,Liu:2019zoy,Brambilla:2019esw}. Understanding the structures and the decay properties of the exotic states is an important and long term issue in the particle physics. In fact, as discussed in Ref.~\cite{Guo:2017jvc}, most of the exotic states are located nearby the thresholds of certain channels and show up clearly molecular nature. A further study was given in Ref.~\cite{Dong:2020hxe} to reveal this mysterious feature using a nonrelativistic effective field theory with open channels, and more investigations can be found in Ref.~\cite{Dong:2021juy} where the spectrum of hadronic molecules in the charm sector were systematically researched with the single channel Bethe-Salpeter equation. 

Even though more and more new resonances in recent experiments need to understand their nature, there are also some states found in early stage, such as $f_0(980)$, $a_0(980)$, $\Lambda(1405)$, and so on,  to be still under debate for their structures and other properties. As we already known, the $\Lambda(1405)$ resonance was discovered earlier in 1961 in the invariant mass spectrum of $\pi \Sigma$ below the $K^- p$ threshold~\cite{Alston:1961zzd}, which was theoretically predicted in Refs.~\cite{Dalitz:1959dn, Dalitz:1960du} as a quasi-bound molecular state existed in the $\bar{K}N$ interaction. The $\Lambda(1405)$ state has been studied extensively in theory after its discovery by the experiments. At the beginning, considered it as the normal three-quark baryon, the constituent quark models failed to reproduce its mass and describe its nature~\cite{Isgur:1978xj,Capstick:1986bm}. Treated as the five-quark baryon, a $\Lambda^*$ state with the mass about 1400 MeV was obtained using the MIT bag model~\cite{Strottman:1979qu}, but its assignment to the $\Lambda(1405)$ resonance was still questionable due to the wave function having other important components except for the part of $\bar{K} N$ and the small ratio of the coupling constants for the channels of $\bar{K} N$ and $\pi \Sigma$.  Using Jaffe and Wilczek's diquark model, it was possible to interpret the $\Lambda(1405)$ state as $J^P = \frac{1}{2}^-$ pentaquark state~\cite{Zhang:2004xt}. In the Skyrme model, the $\Lambda(1405)$ state was explained as the kaon bound state~\cite{Callan:1985hy}. Constructed an effective potential from the chiral meson-baryon Lagrangian and using the coupled channel Lippmann-Schwinger equation, the $\Lambda(1405)$ resonance was reproduced as quasi-bound state of $\bar{K}N$ and the experimental data was well described in Refs.~\cite{Kaiser:1995eg,Kaiser:1996js}. Furthermore, also with the coupled channel Lippmann-Schwinger equation (later denoted as the coupled channel Bethe-Salpeter equation under the on-shell description~\cite{Oller:2000ma}), but derived the interaction potential only from the lowest order chiral Lagrangian, Ref. \cite{Oset:1997it} also dynamically generated the $\Lambda(1405)$ resonance with one cutoff for the loop functions, and also described the experimental data well. 

Furthermore, using the chiral unitary approach (ChUA)~\cite{Oset:1997it,Oller:2000ma}, where more details about the ChUA can be referred to the recent reviews \cite{Oller:2019opk,MartinezTorres:2020hus,Oller:2020guq,Guo:2020hli}, more further investigations had been done in Refs.~\cite{Oller:2000fj,Oset:2001cn,Jido:2002yz,GarciaRecio:2002td,Jido:2003cb,GarciaRecio:2003ks,Hyodo:2007jq,Hyodo:2008xr} for the interactions of $\bar{K}N$ and its coupled channels to reveal the nature of the $\Lambda(1405)$ state. Especially, two-poles structure was found in different Riemann sheets for the $\Lambda(1405)$ state in Ref~\cite{Oller:2000fj}, which was confirmed in the later work of~\cite{Jido:2002yz,GarciaRecio:2002td} and investigated with more details about the nature of two poles in Ref.~\cite{Jido:2003cb}. In Ref.~\cite{GarciaRecio:2003ks} the quark mass dependence for these poles was discussed. The pole structure of the $\Lambda(1405)$ state was looked inside with the interactions of coupled channels and single channel in Ref.~\cite{Hyodo:2007jq}. In fact, the two-poles structure of the $\Lambda(1405)$ resonance was predicted for the first time in Ref.~\cite{Fink:1989uk} using the coupled channel Lippmann-Schwinger equation where the cloudy quark bag model was taken for the potentials. The progress on the ChUA for the $\Lambda(1405)$ resonance can be found in the reviews~\cite{Hyodo:2011ur}. 

Note that, Refs.~\cite{Kaiser:1995eg,Kaiser:1996js} had taken into account the next leading order of chiral Lagrangian to extract the potential. And the direct and crossed Born terms were considered to the interaction potentials in Ref.~\cite{Oller:2000fj}. Moreover, with the ChUA, Ref.~\cite{Borasoy:2005ie} considering both the leading order contact (Weinberg-Tomozawa) term, the direct and crossed Born terms (the $s$- and $u$-channel contributions~\cite{Xiao:2016ogq}), and the next leading order (NLO) contact terms to derived the interaction potentials, where the contributions to the fitting results were discussed in details and they found that the ones for the Born terms were small, for more investigations see Refs.~\cite{Oller:2005ig,Oller:2006jw,Borasoy:2006sr,Ikeda:2011pi,Ikeda:2012au,Guo:2012vv,Ramos:2016odk}. Following Ref.~\cite{Bruns:2010sv}, a Bethe-Salpeter framework with full off-shell dependence was  adopted in Ref.~\cite{Mai:2012dt} for the interactions of $\bar{K}N$ and its coupled channels, where the two-pole structure was confirmed and the off-shell effect was impacted slightly the description of experimental data. A further investigation was given in Ref.~\cite{Mai:2014xna} with more constraints from the new experimental data. Based on the results of lattice QCD and the ChUA in the finite volume, Ref.~\cite{Meissner:2020khl} argued that the two-pole structure was a fact of QCD and suggested that the two-pole structure should be put in the main listings of Particle Data Group (PDG)~\cite{Zyla:2020zbs}. To view the origin of these poles, a comparative analysis of different theoretical models was performed in Ref.~\cite{Cieply:2016jby}. The wave function and the compositeness for the $\Lambda(1405)$ resonance  were discussed in details in Refs.~\cite{Sekihara:2014kya,Kamiya:2016jqc} to address the critical issues of its molecular picture. Using the ChUA and a full analysis of low-energy experimental data including the differential cross section data for the first time, Ref.~\cite{Sadasivan:2018jig} determined the two poles of $\Lambda(1405)$ by a simultaneous analysis of $s$- and $p$-waves of the scattering amplitude. With more constraints from the vector-baryon interactions in the ChUA, the two poles was well reproduced in Ref.~\cite{Khemchandani:2018amu}, where one new state of isospin $I=1$ was claimed. In a chiral quark model, it was also supported that the $\Lambda(1405)$ state was a $\bar{K}N$ bound state in Ref.~\cite{Qin:2020gxr} and had the two-pole structure. Whereas, in the Skyrme model, the $\Lambda(1405)$ state was assumed to be a narrow $\bar{K}N$ Feshbach resonance in Ref.~\cite{Ezoe:2020piq}. Using a model-independent of the Uniformized Mittag-Leffler expansion, Ref.~\cite{Yamada:2021cjo} fitted the experimental data well and supported the  single-pole picture of  the $\Lambda(1405)$.  Recently, also with the full off-shell dependence for the effective potential, the two-pole structure was confirmed once again in Ref.~\cite{Ren:2021yxc} by applying subtractive renormalization for the coupled channel scattering amplitudes. Besides, there were also some experimental results for the two-pole structure of the $\Lambda(1405)$ state~\cite{Zychor:2007gf,Lu:2013nza}. More about the status of the $\Lambda(1405)$ resonance can be found in the review of PDG~\cite{Zyla:2020zbs} by Mei{\ss}ner and Hyodo, and in the recent reviews of~\cite{Mai:2020ltx,Hyodo:2020czb}.

In the present work, to understand more about the properties of $\Lambda(1405)$, we revisit the coupled channel interactions of $\bar{K}N$ and its coupled channels with the ChUA following the method of Ref.~\cite{Ahmed:2020kmp}, where the pole structures of the $\sigma$ or called $f_0(500)$, $f_0(980)$ and $a_0(980)$ were investigated by their corresponding compositeness, wave functions and radii in the coupled and single channel interactions. Our work is organized as follows. In the next section, we briefly introduce the formalism of coupled channel interaction within the ChUA, where Bethe-Salpeter equation and the meson-baryon interaction potentials are provided. Furthermore, we discuss the evaluations of the couplings, the compositeness, the wave functions and the radii for two poles of $\Lambda(1405)$. In the following section, we show our results of the coupled and single channel interactions. At the end, it is our conclusions.

\section{Formalism}

In this section, we first revisit the formalism of the ChUA for the meson-baryon interactions, where the interaction potentials are derived from the lowest order chiral Lagrangian, and the scattering amplitudes are evaluated with the coupled channel Bethe-Salpeter equation. Next, we introduce the calculations of the couplings, the wave functions, the compositeness and the radii of the resonances by extending the work of~\cite{Ahmed:2020kmp} to the baryonic sector.

\subsection{S-wave scattering amplitude in coupled channels}

The lowest order baryonic Lagrangian that describes the low-energy interaction of pseudoscalar mesons with baryon fields was discussed in details in Refs.~\cite{Meissner:1993ah,Pich:1995bw,Ecker:1994gg,Bernard:1995dp}, where the interaction Lagrangian at the lowest order momentum comes from the $\Gamma_{\mu}$ term in the covariant derivative, given by 
\begin{equation}
	L_{1}^{(B)}=\left\langle\bar{B} i \gamma^{\mu} \frac{1}{4 f^{2}}\left[\left(\Phi \partial_{\mu} \Phi-\partial_{\mu} \Phi \Phi\right) B-B\left(\Phi \partial_{\mu} \Phi-\partial_{\mu} \Phi \Phi\right)\right]\right\rangle.
	\label{eq:lagrangian}
\end{equation}
where the symbol $\langle\rangle$ stands for the trace of $SU(3)$ matrices, and $SU(3)$ matrices for the pseudoscalar mesons and the lowest-lying $\frac{1}{2}^{+}$ baryon octet are written as
\begin{equation}
\Phi=\left(\begin{array}{ccc}
\frac{1}{\sqrt{2}} \pi^{0}+\frac{1}{\sqrt{6}} \eta & \pi^{+} & K^{+} \\
\pi^{-} & -\frac{1}{\sqrt{2}} \pi^{0}+\frac{1}{\sqrt{6}} \eta & K^{0} \\
K^{-} & \bar{K}^{0} & -\frac{2}{\sqrt{6}} \eta
\end{array}\right),
\end{equation}

\begin{equation}
	B=\left(\begin{array}{ccc}
		\frac{1}{\sqrt{2}} \Sigma^{0}+\frac{1}{\sqrt{6}} \Lambda & \Sigma^{+} & p \\
		\Sigma^{-} & -\frac{1}{\sqrt{2}} \Sigma^{0}+\frac{1}{\sqrt{6}} \Lambda & n \\
		\Xi^{-} & \Xi^{0} & -\frac{2}{\sqrt{6}} \Lambda
	\end{array}\right).
\end{equation}

Then, from the lowest order of chiral Lagrangian in Eq.~\eqref{eq:lagrangian}, we obtain the $s$-wave interaction potentials for the coupled channels of $\bar{K}N$. There are four coupled channels in isospin $I=0$ and strangness $S=-1$ sector,  $\bar{K}N$, $\pi\Sigma$, $\eta\Lambda$ and $K\Xi$, denoted as channel 1, 2, 3 and 4, respectively. Finally, we get the interaction potentials as follow~\cite{Jido:2003cb},
\begin{equation}
	V_{i j}=-C_{i j} \frac{1}{4 f^{2}}\left(2 \sqrt{s}-M_{i}-M_{j}\right)\left(\frac{M_{i}+E}{2 M_{i}}\right)^{1 / 2}\left(\frac{M_{j}+E^{\prime}}{2 M_{j}}\right)^{1 / 2},
	\label{eq:Vij}
\end{equation}
where $E$, $E^{\prime}$ are the energies of the initial and final mesons, $M_{i}$, $M_{j}$ the masses of the initial and final baryons, and the coefficient matrix $C_{ij}$ is symmetric, given in Ref.~\cite{Oset:1997it}. For the meson decay constant, we take $f=1.123f_\pi$~\cite{Jido:2003cb}, where the one $f_\pi$ is the weak pion decay constant, having $f_\pi=93$ MeV~\cite{Oset:1997it}.

The scattering amplitudes of the coupled channels ($T$) can be solved by Bethe-Salpeter equations with the on-shell description~\cite{Oset:1997it},
\begin{equation}
	T=[1-V G]^{-1}V,
	\label{eq:T}
\end{equation}
where the matrix $V$ is constructed by Eq.~\eqref{eq:Vij}. The element of diagonal matrix $G$ the loop function of  intermediate mesons and baryons in the $l$-th channel is given by
\begin{equation}
	G_{l} =i \int \frac{d^{4} q}{(2 \pi)^{4}} \frac{2 M_{l}}{(P-q)^{2} - M_{l}^{2}+i \epsilon} \frac{1}{q^{2}-m_{l}^{2}+i \epsilon}\text{ ,}
\end{equation}
where $P =p_{1}+k_{1}$ is the total momentum of the meson-baryon system. Note that the $G$ function is logarithmically divergent, which need to be regularized. Using the three-momentum cutoff (CO) method, one can obtain~\cite{Oset:1997it},
\begin{equation}
	G_{l}(s)=\int_{0}^{q_{max}} \frac{q^{2} d q}{2 \pi^{2}} \frac{1}{2 \omega_{l}(q)} \frac{M_{l}}{E_{l}(q)} \frac{1}{p^{0}+k^{0}-\omega_{l}(q)-E_{l}(q)+i \epsilon}, 
	\label{eq:gqcm}
\end{equation}
where $\omega_{l}(q)=\sqrt{q^2+m_{l}^2}$, $E_{l}(q)=\sqrt{q^2+M_{l}^2}$, and $p^{0}+k^{0}=\sqrt{s}$, $q_{max}$ is the only free parameter.
On the other hand, with the dimensional regularization (DR) method, we have the one from Ref.~\cite{Oller:2000fj},
\begin{equation}
	\begin{aligned}
	G_{l}(s)=& \frac{2 M_{l}}{16 \pi^{2}}\left\{a_{\mu}+\ln \frac{M_{l}^{2}}{\mu^{2}}+\frac{m_{l}^{2}-M_{l}^{2}+s}{2 s} \ln \frac{m_{l}^{2}}{M_{l}^{2}}\right.\\
		&+\frac{q_{cml}(s)}{\sqrt{s}}\left[\ln \left(s-\left(M_{l}^{2}-m_{l}^{2}\right)+2 q_{cml}(s) \sqrt{s}\right)\right.\\
		&+\ln \left(s+\left(M_{l}^{2}-m_{l}^{2}\right)+2 q_{cml}(s) \sqrt{s}\right) \\
		&-\ln \left(-s-\left(M_{l}^{2}-m_{l}^{2}\right)+2 q_{cml}(s) \sqrt{s}\right) \\
		&\left.\left.-\ln \left(-s+\left(M_{l}^{2}-m_{l}^{2}\right)+2 q_{cml}(s) \sqrt{s}\right)\right]\right\},
	\end{aligned}
	\label{eq:gdr}
\end{equation}
where $\mu$ is the regularization scale, $a_{\mu}$ the subtraction constant, and $q_{cml}(s)$ the three momentum of the particles in the center-of-mass (CM) frame, given by 
\begin{equation}
	q_{cml}(s)=\frac{\lambda^{1 / 2}\left(s, M_{l}^{2}, m_{l}^{2}\right)}{2 \sqrt{s}},
\end{equation}
with the usual K\"allen triangle function $\lambda(a, b, c)=a^{2}+b^{2}+c^{2}-2(a b+a c+b c)$.

In the ChUA, the mass ($M_{r}$) and decay width ($\Gamma_{r}$) of the resonance for the coupled channels interactions can be determined from the pole ($s_{p}$) in the general second Riemann sheets, having $\sqrt{s_{p}}=M_{r} - i \Gamma_{r}/2$ 
\footnote{Note that for the case of a bound state without open (coupled) channel to decay, the pole will be located in the first Riemann and the real axes, and thus, the pole has no width, written $\sqrt{s_{p}}=M_{r}$.}. 
Thus, to search for the pole in the second Riemann sheet, the scattering amplitude should be extrapolated to the complex plane, which can be fulfilled by the $G_{l}(s)$ function defined to the second Riemann sheet,
\begin{equation}
	G_{l}^{(I I)}(s)=G_{l}^{(I)}(s)-2 i \operatorname{Im} G_{l}^{(I)}(s)=G_{l}^{(I)}(s)+\frac{i}{2 \pi} \frac{M_{l}q_{cml}(s)}{\sqrt{s}},
	\label{eq:GII}
\end{equation}
where $G_{l}^{(I)}$ corresponds to the propagator function in the first Riemann sheet, see Eq.~\eqref{eq:gqcm} or Eq.~\eqref{eq:gdr}.

\subsection{Definition of the couplings and wave functions}

To probe more structure properties of a resonance, we make a further investigation on its couplings to the other coupled channels, the wave functions, and so on. For the case of the meson-baryon interaction, the couplings can be defined by the residue of the pole,
\begin{equation}
	g_{i}^{2}=\lim _{\sqrt{s} \rightarrow \sqrt{s_{p}}}\left(\sqrt{s}-\sqrt{s_{p}}\right) T_{i i}.
\end{equation}

Furthermore, we can generalize Weinberg’s rule~\cite{Weinberg:1965zz} for the bound state or resonance to the formalism of the ChUA with the couplings obtained~\cite{Aceti:2012dd},
\begin{equation}
	-\sum_{i} g_{i}^{2}\left[\frac{d G_{l}}{d \sqrt{s}}\right]_{\sqrt{s}=\sqrt{s_{p}}}=1,
\end{equation}
which is valid only for the case of a pure molecular state. However, in some cases, a bound state or resonance contains not only the molecular parts, but also some other non-molecular components. Thus, for the case of the composite state, this sum rule can be generalized as,
\begin{equation}
	-\sum_{i} g_{i}^{2}\left[\frac{d G_{l}}{d \sqrt{s}}\right]_{\sqrt{s}=\sqrt{s_{p}}}=1-Z,
	\label{eq:components}
\end{equation}
where $Z$ represents the probability of the other non-molecular components contained in the bound state or resonance. Note that for a specified channel, the $G_l$ function should be extrapolated to the right Riemann sheet for a corresponding pole. More details about the sum rule can be referred to Refs.~\cite{Baru:2003qq,Sekihara:2012xp,Xiao:2012vv,Hyodo:2013iga,Aceti:2014ala,Guo:2015daa}

In order to make further study on the components and sources of the states, we construct the wave function of resonances in a small distances. As done in Ref.~\cite{YamagataSekihara:2010pj}, the wave function of a resonance in coordinate space was given by
\begin{equation}
	\phi(\vec{r})=\int_{q_{\max }} \frac{d^{3} \vec{p}}{(2 \pi)^{3 / 2}} e^{i \vec{p} \cdot \vec{r}}\langle\vec{p} \mid \Psi\rangle.
\end{equation}
Performing the angle integration of the momentum, one can have~\cite{Ozpineci:2013qza}
\begin{equation}
	\begin{aligned}
		\phi(\vec{r})=\frac{1}{(2 \pi)^{3 / 2}} \frac{4 \pi}{r} \frac{1}{C} \int_{q_{\max }} p d p \sin (p r)\times \frac{\Theta\left(q_{\max }-|\vec{p}|\right)}{E-\omega_{1}(\vec{p})-\omega_{2}(\vec{p})}.
	\end{aligned}
\end{equation}
where $C$ is the normalization constant, and $E \equiv \sqrt{s_{p}}$. From the wave function obtained, we can calculate the form factor,
\begin{equation}
	\begin{aligned}
		F(\vec{q}) &=\int d^{3} \vec{r} \phi(\vec{r}) \phi^{*}(\vec{r}) e^{-i \vec{q} ^ {\prime} \cdot \vec{r}} \\
		&=\int d^{3} \vec{p}\times \frac{\theta(\Lambda-p) \theta(\Lambda-|\vec{p}-\vec{q}|)}{\left[E-\omega_{1}(p)-\omega_{2}(p)\right]\left[E-\omega_{1}(\vec{p}-\vec{q})-\omega_{2}(\vec{p}-\vec{q})\right]},
	\end{aligned}
\end{equation}
with a normalization to keep $F(q=0) \equiv 1$ and $\theta(\Lambda-p)$ is the step function. Finally, we can get the radius of the resonance with the form factor at origin,
\begin{equation}
	\left\langle r^{2}\right\rangle=-6\left[\frac{\mathrm{d} F(q)}{\mathrm{d} q^{2}}\right]_{q^{2}=0},
	\label{eq:radii}
\end{equation}
which is consistent with the one obtained from the tail of the wave functions~\cite{Sekihara:2012xp} for the state not close to the threshold, see more discussions in Ref.~\cite{Ahmed:2020kmp}.

\section{Results}

We first revisit the $\bar{K}N$ interactions with coupled channels of $\bar{K}N$, $\pi\Sigma$, $\eta\Lambda$ and $K\Xi$ in the isospin basis as done in Ref. \cite{Oset:1997it}. Since the two relevant poles found in the second Riemann sheet (shown below) couple strongly to the $\bar{K}N$ and $\pi\Sigma$ channels, we pay more attention to these two channels. Then we research the interactions with only two coupled channels of $\bar{K}N$ and $\pi\Sigma$, and the interactions of these two singled channels in the isospin $I=0$ and strangeness $S=-1$ sector. The $\Lambda(1405)$ resonance is dynamically generated in the coupled channels interactions, where two-pole structure is found. In order to make further understanding of the components and properties for this resonance, and reveal the nature of two poles, we investigate the couplings, the compositeness, the wave functions and the radii of two poles for this state in the coupled channel and single channel interactions. 

\subsection{Coupled channels approach}

\begin{figure}
		\centering
		\includegraphics[width=0.6\linewidth]{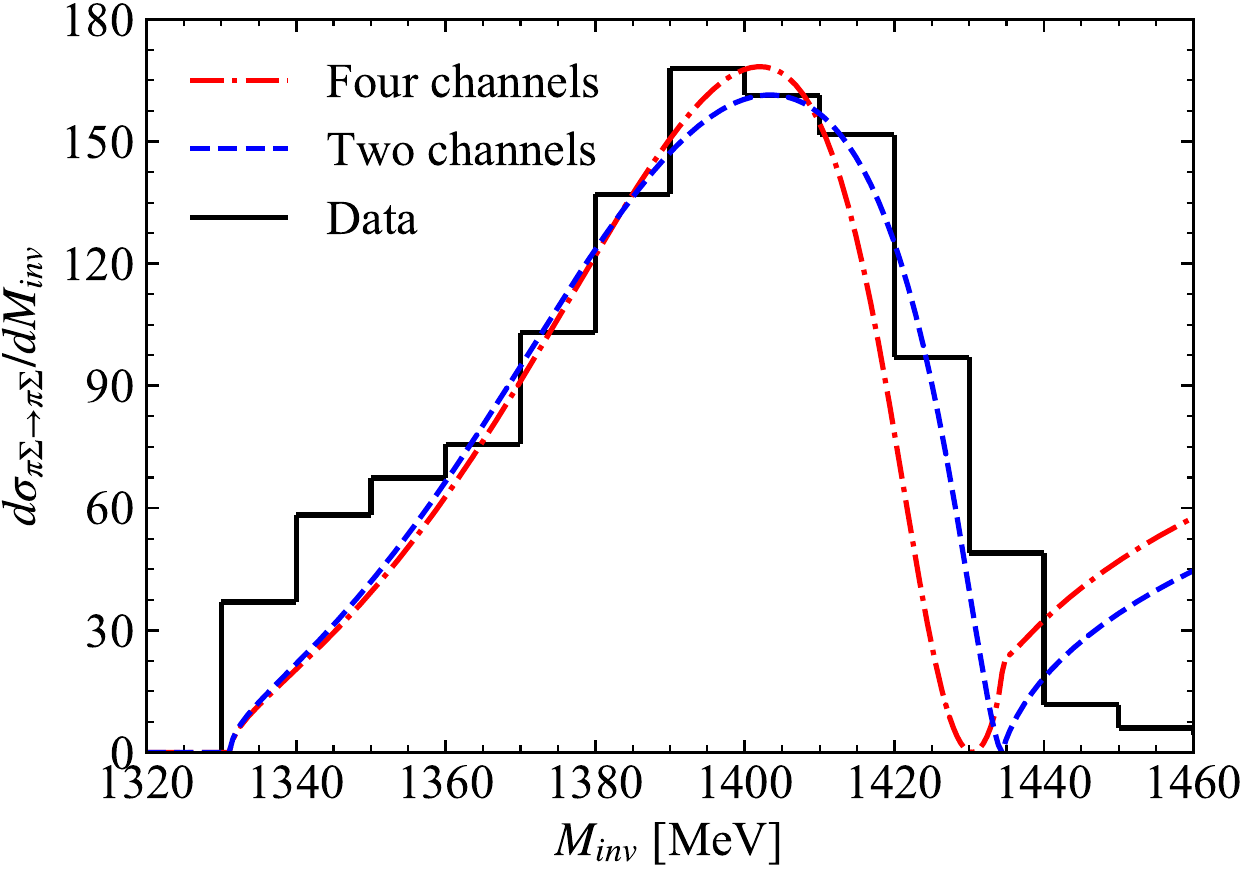}
	\caption{$\pi\Sigma$ invariant mass spectra of $\pi \Sigma \rightarrow \pi \Sigma$ with a normalization coefficient $C = 113$ for four coupled channels and $C = 110$ for two coupled channels. Data were taken from Ref. \cite{Hemingway:1984pz}.}
	\label{fig:42Gamma}
\end{figure} 

\begin{figure}
  \centering
  \includegraphics[width=0.49\linewidth]{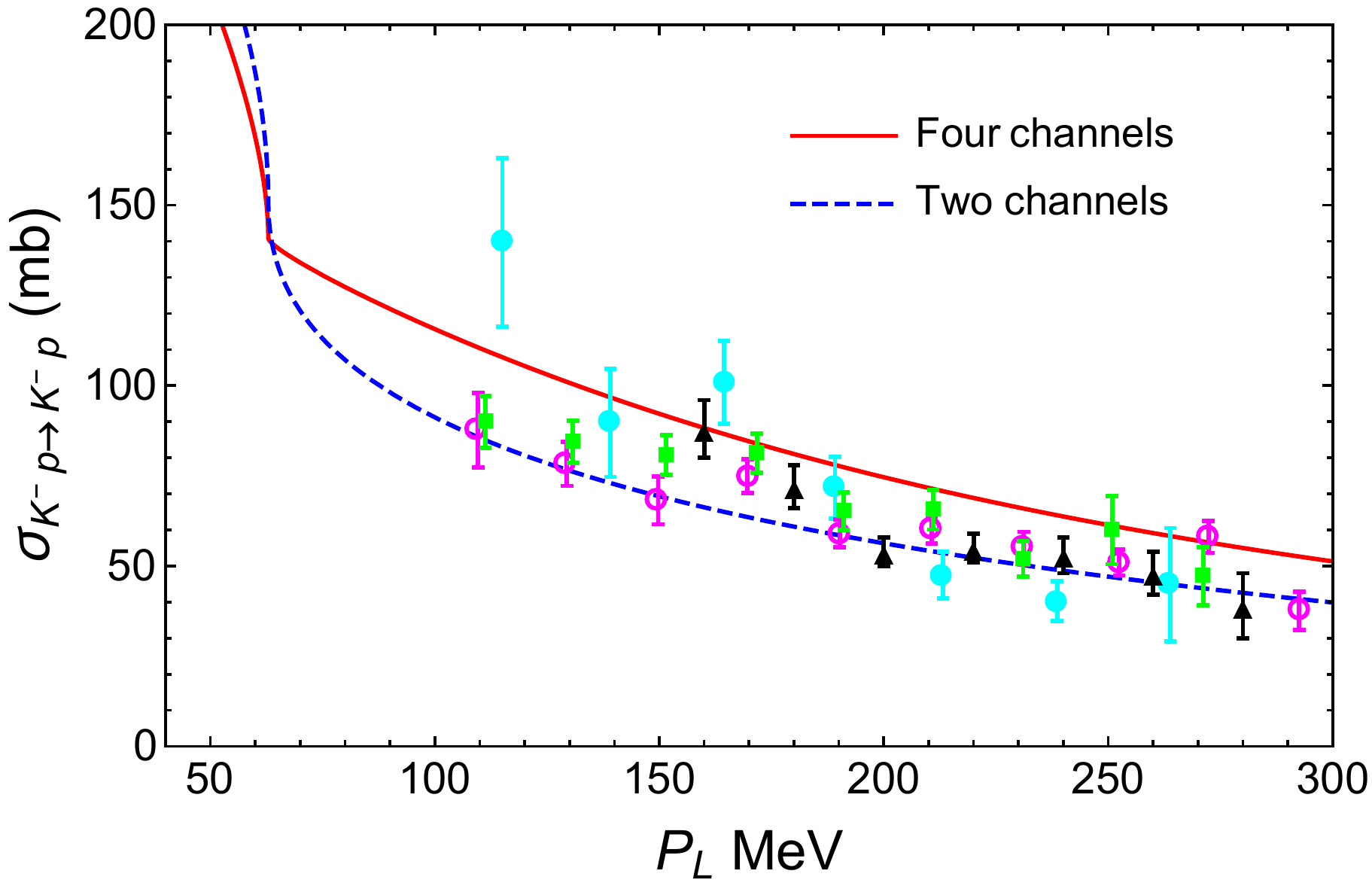} 
  \includegraphics[width=0.49\linewidth]{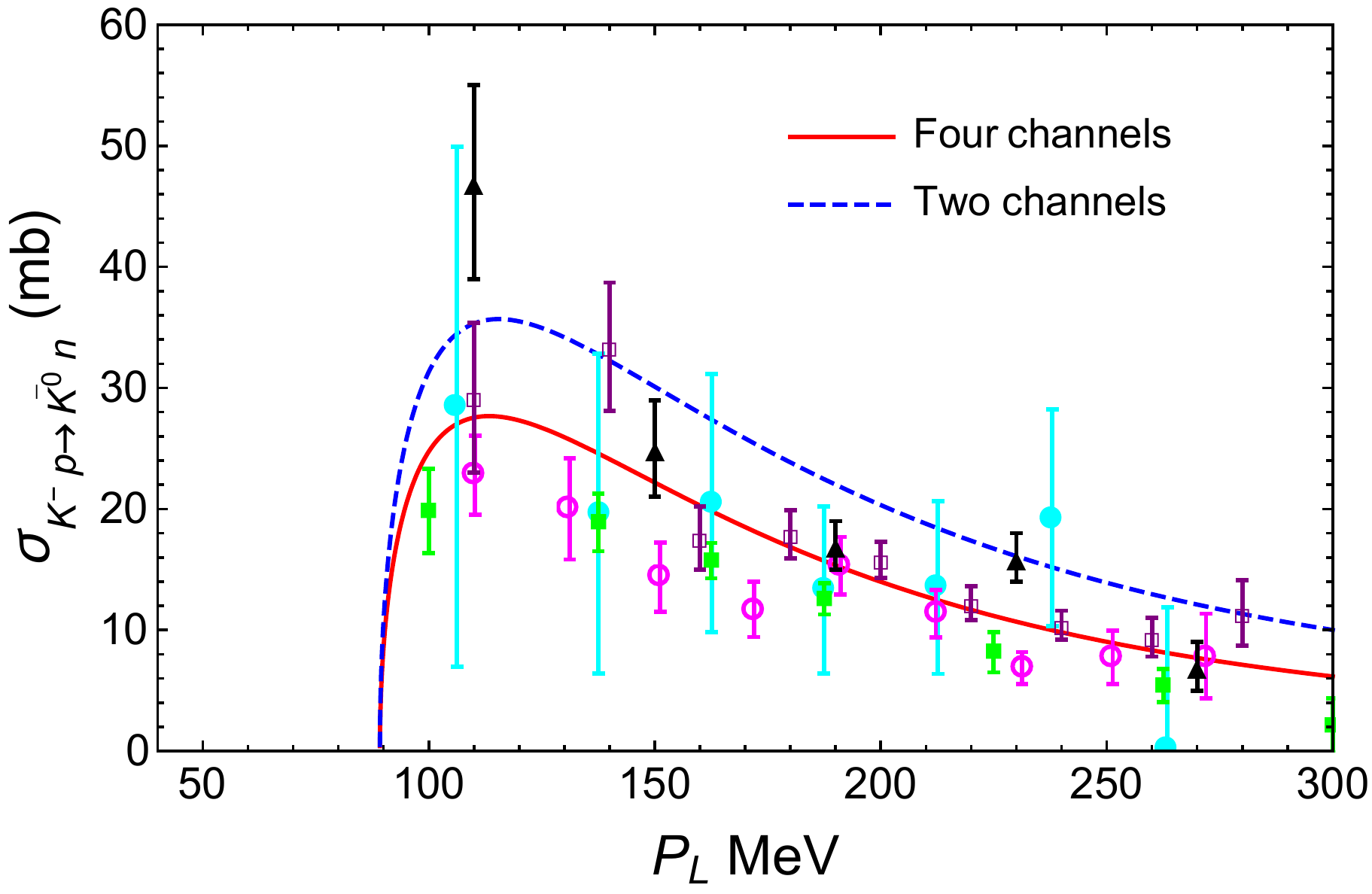} 
  \includegraphics[width=0.49\linewidth]{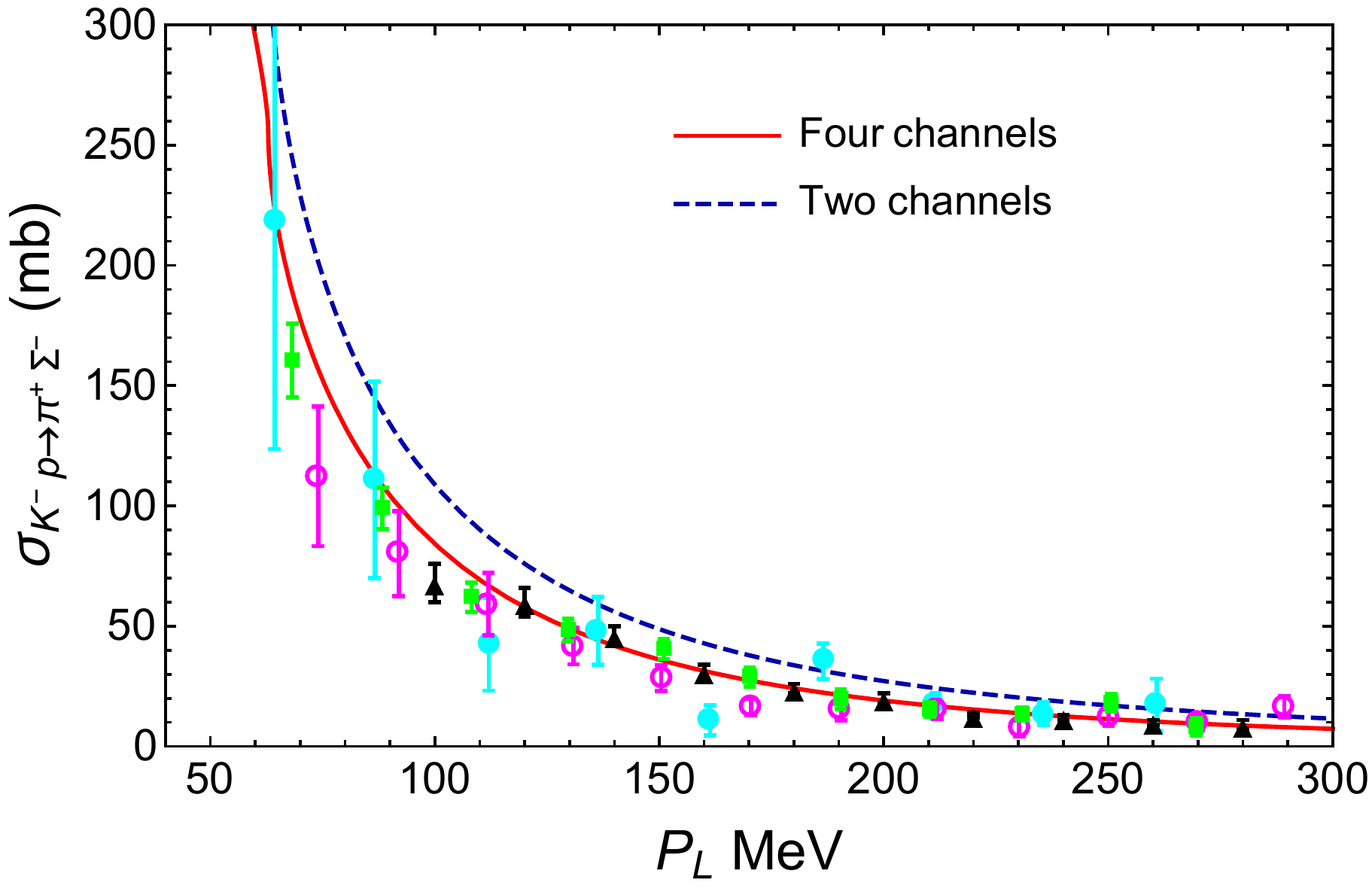} 
  \includegraphics[width=0.49\linewidth]{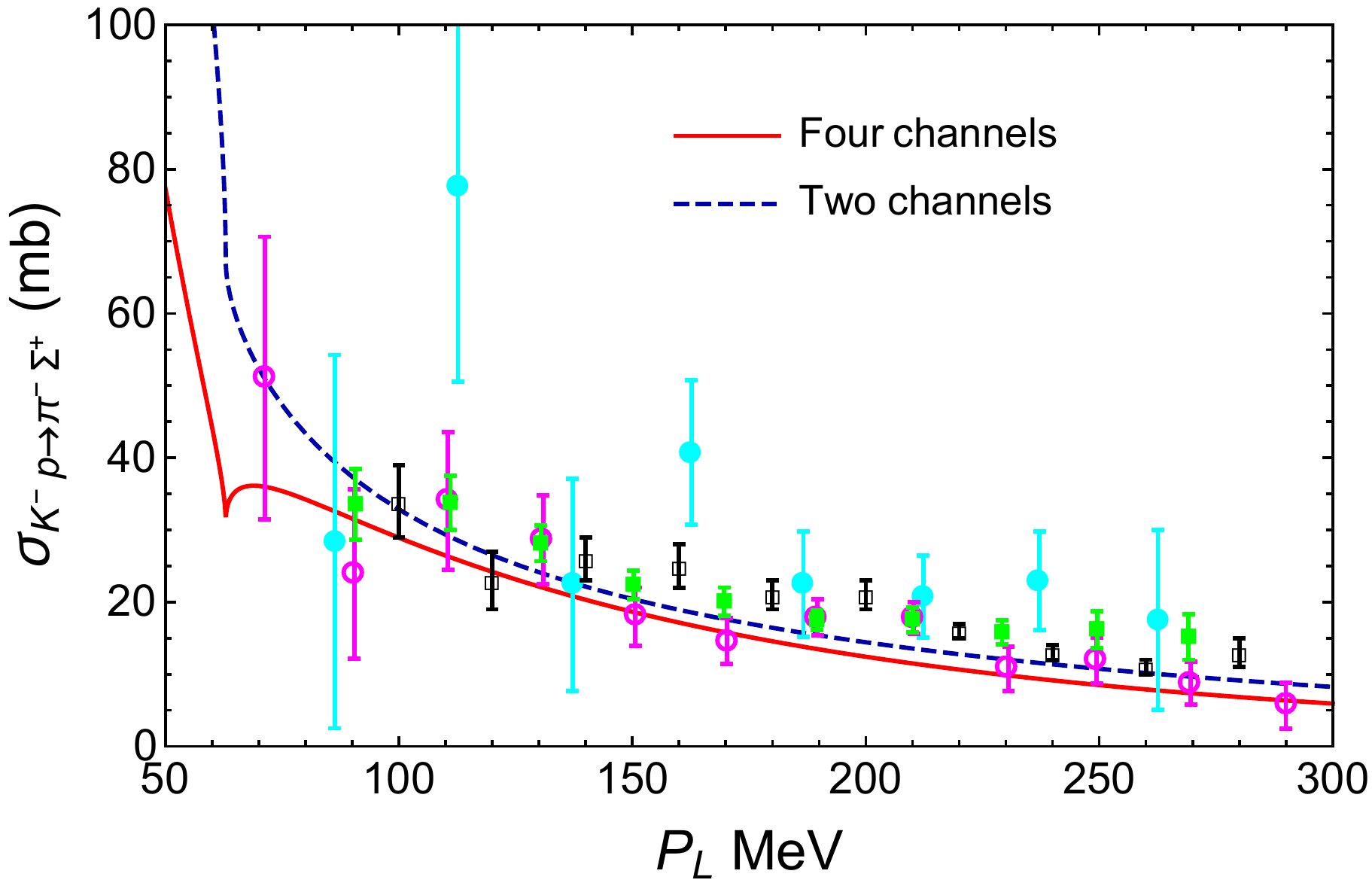}     
  \includegraphics[width=0.49\linewidth]{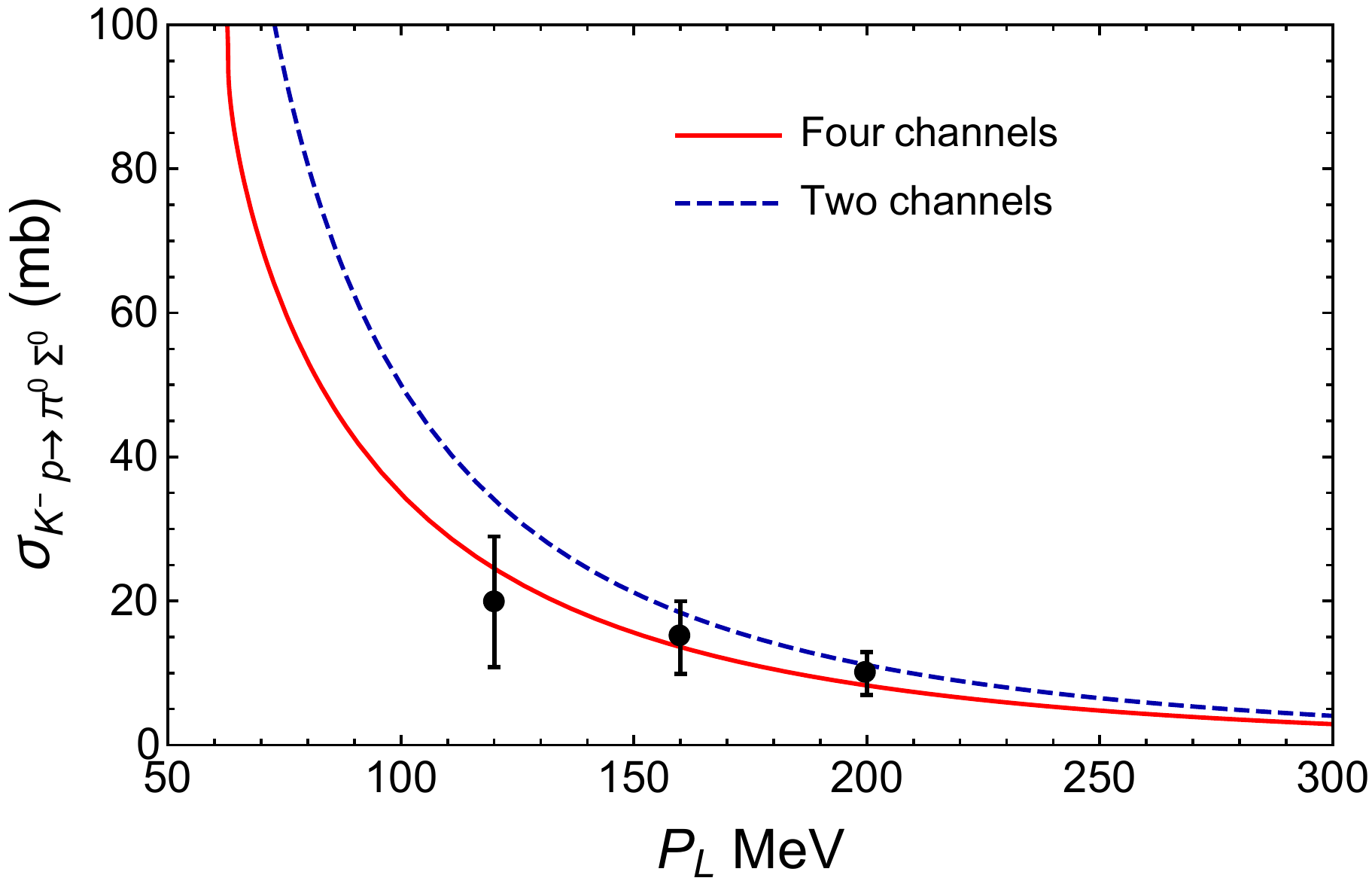}  
  \caption{Cross sections for $K^{-}p $ $\to$ $ K^{-}p$, $\bar{K^{0}}n$, $\pi^{+} \Sigma^{-}$, $\pi^{-} \Sigma^{+}$ , and $\pi^{0} \Sigma^{0}$ in the lab frame. Solid (Red) line: results of four coupled channels. Dashed (blue) line: results of two coupled channels. See Refs.~\cite{Humphrey:1962zz,Sakitt:1965kh,Kim:1967zze,Kittel:1966zz,Ciborowski:1982et,Evans:1983hz,Kim1966,Martin:1980qe} for the details of the experimental data.}
\label{fig:sigma}
\end{figure}

First, we show our results for the coupled channel interactions with four coupled channels $\bar{K}N$, $\pi\Sigma$, $\eta\Lambda$, $K\Xi$, and two coupled channels $\bar{K}N$, $\pi\Sigma$. As discussed in the last section, the free parameters of the ChUA are $q_{max}$ for the cutoff method, or the regularization scale $\mu$ and subtraction constant $a_{\mu}$ for the dimensional regularization method, which are used to regularize the propagator functions. The values of these parameters for the case of four coupled channels are $\mu=q_{max}=630$ MeV~\cite{Oset:1997it} and $a_{\bar{K} N}=-1.84$, $a_{\pi \Sigma}=-2.00$, $a_{\eta \Lambda}=-2.25$, $ a_{K \Xi}=-2.67$, which were taken from Ref.~\cite{Jido:2003cb}. For the case of two coupled channels, we obtained $\mu=q_{max}=623$ MeV and $a_{\bar{K} N}=-1.86$, $a_{\pi \Sigma}=-2.09$ by fitting the $\pi\Sigma$ invariant mass distributions, see our results of Fig.~\ref{fig:42Gamma}, where the results of both cases described the experimental data well. In Fig.~\ref{fig:42Gamma}, we show the results of the $\pi\Sigma$ invariant mass distributions, which are given by
\begin{equation}
	\frac{d \sigma}{d M_{inv}}=C\left|T_{\pi \Sigma \rightarrow \pi \Sigma}\right|^{2} k_{\pi},
\end{equation}
where $C$ is a normalized constant, $T_{\pi \Sigma \rightarrow \pi \Sigma}$ the scattering amplitude of $\pi \Sigma$ channel obtained from Eq.~\eqref{eq:T} for $I=0$ sector, and $k_{\pi}$ the $\pi$ momentum in the CM frame of $\pi \Sigma$ system. For the case of four coupled channel, the results of Fig.~\ref{fig:42Gamma} are consistent with the ones in Ref.~\cite{Oset:1997it}. The results for the case of two coupled channel in Fig.~\ref{fig:42Gamma} look like a bit better. There are also some other experimental data as shown in Fig.~\ref{fig:sigma}. The results of Fig.~\ref{fig:sigma} are the low energy total cross sections of $K^{-}p$ scattering to various final states
\footnote{Since we focus on the origin of two poles and their different properties in the present work, we did not include these data for the fit mentioned above in the two coupled channels case. Due to large uncertainties from the data,  for the simplicity we do not show the results of the combined fit with all these data, of which the total chi square is not much different.}, 
which are given by
\begin{equation}
\sigma_{i j}=\frac{1}{4 \pi} \frac{M M^{\prime}}{s} \frac{k^{\prime}}{k}\left|T_{i j}\right|^{2},
\label{eq20}
\end{equation}
where $M$($M^{\prime}$) is the initial(final) baryon mass and $k$($k^{\prime}$) the initial(final) three momentum in the CM frame. It is worth to mention that, since some of the physical decay channels has mixing with isospin $I=0$ and $I=1$ components, we consider the mixing in our calculations, where the details of the amplitudes for $I=1$ are not shown and can be referred to Ref.~\cite{Oset:1997it} for more details. As one can see in Fig.~\ref{fig:sigma}, the results obtained for four and two coupled channels are consistent with the experimental data within the uncertainties.

\begin{figure}
 \centering
		\includegraphics[width=0.49\linewidth]{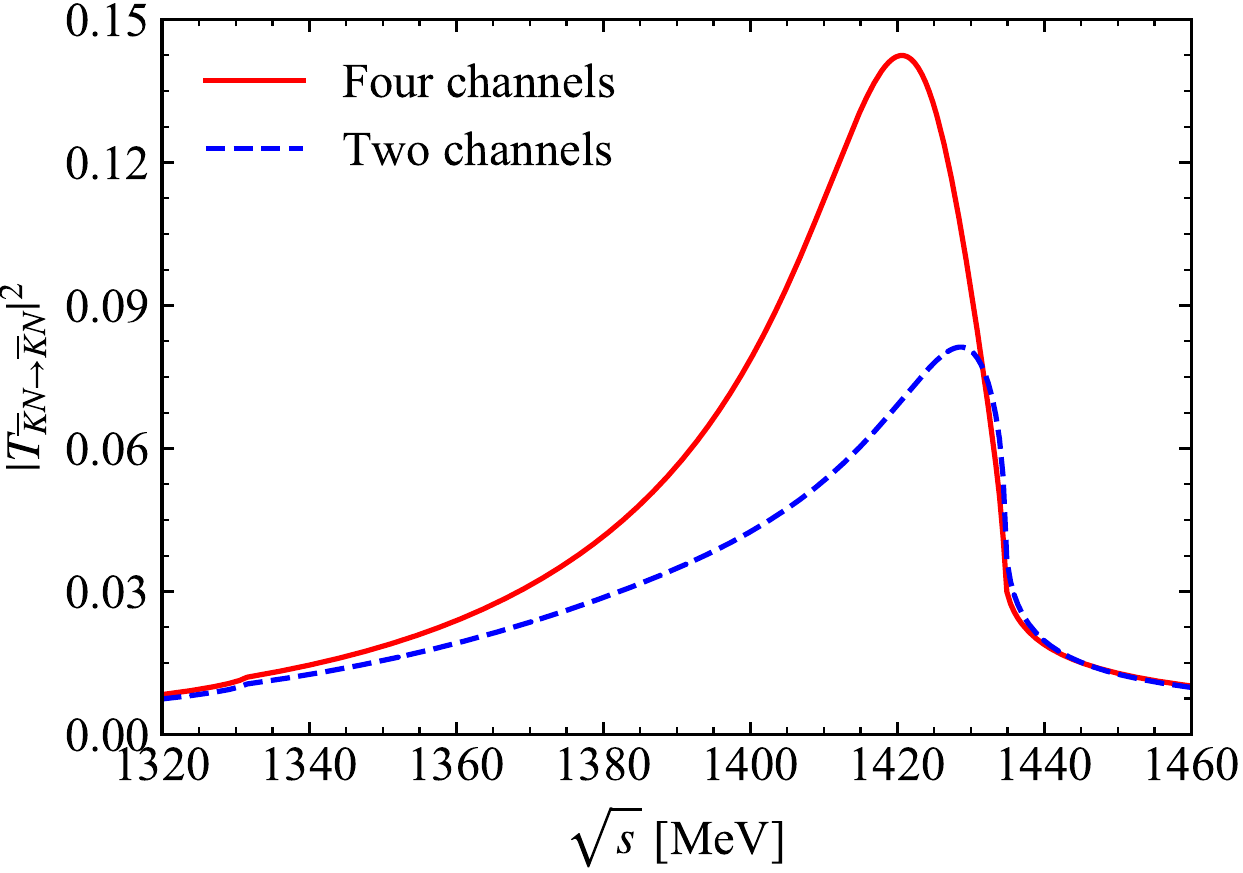} 
		\includegraphics[width=0.49\linewidth]{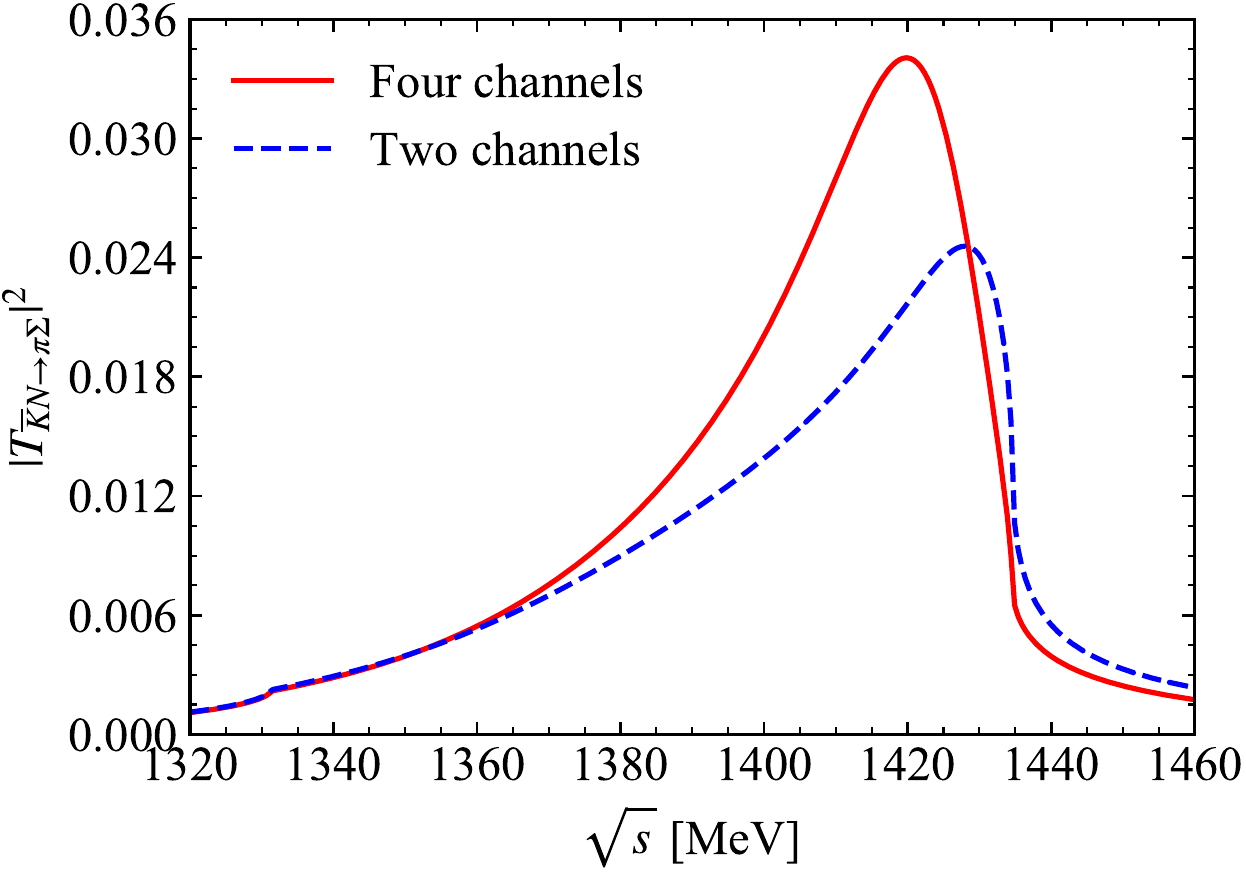} 
		\includegraphics[width=0.49\linewidth]{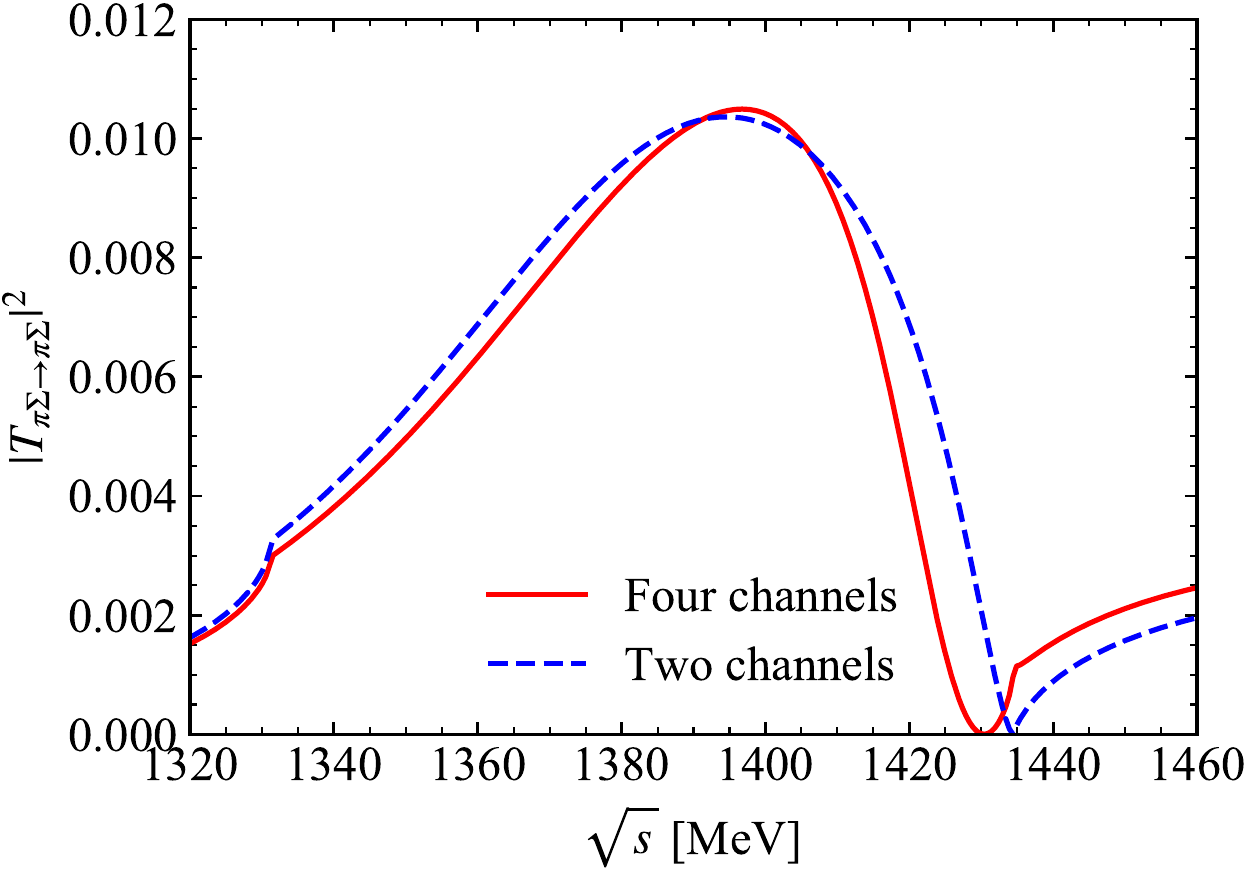}  
	\caption{Modulus square of $\bar{K}N\rightarrow\bar{K}N$, $\bar{K}N\rightarrow\pi\Sigma$ and $\pi\Sigma\rightarrow\pi\Sigma$ amplitudes.}
	\label{fig:42T}
\end{figure}

\begin{figure}
		\centering
		\includegraphics[width=0.49\linewidth]{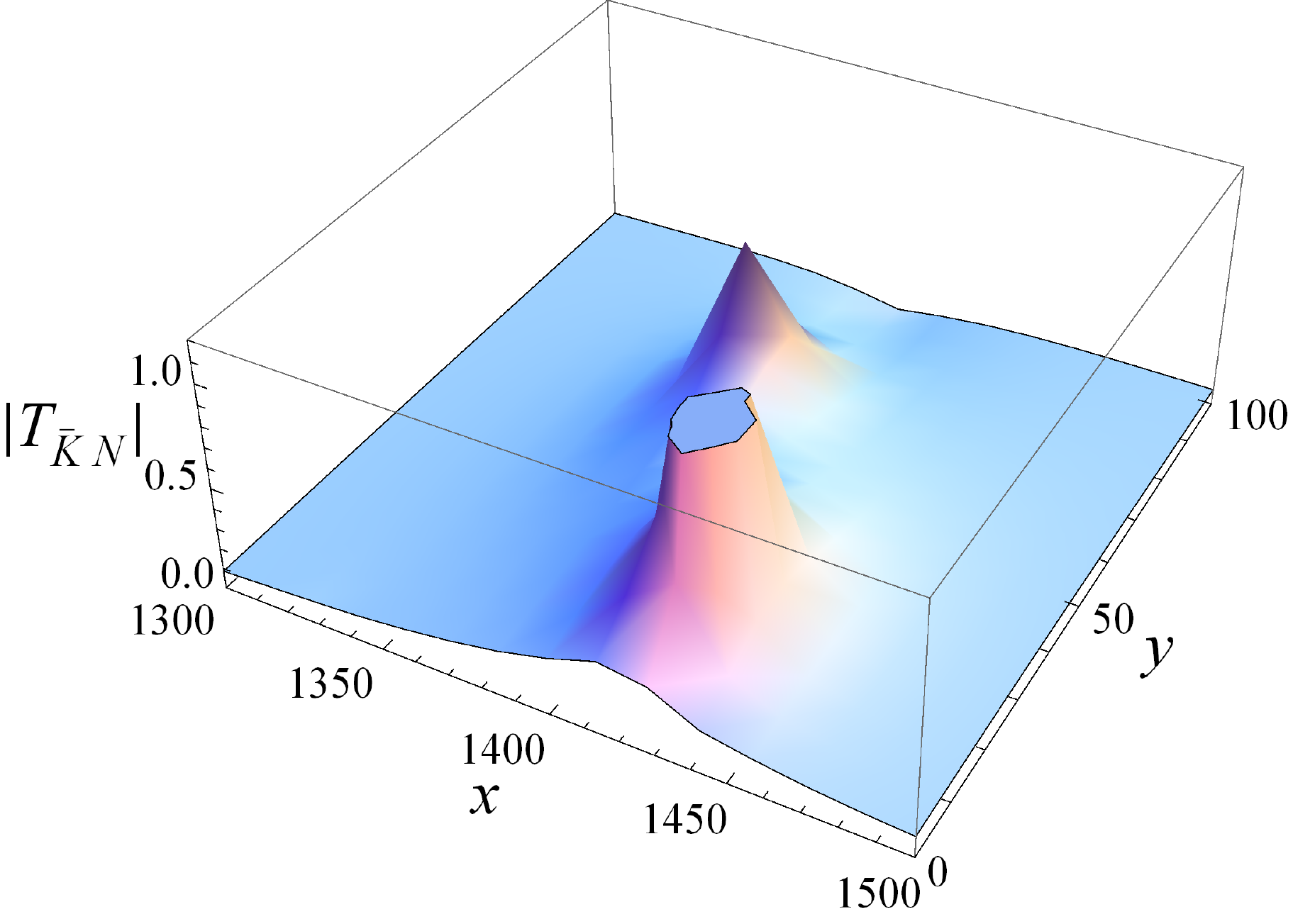} 
		\includegraphics[width=0.49\linewidth]{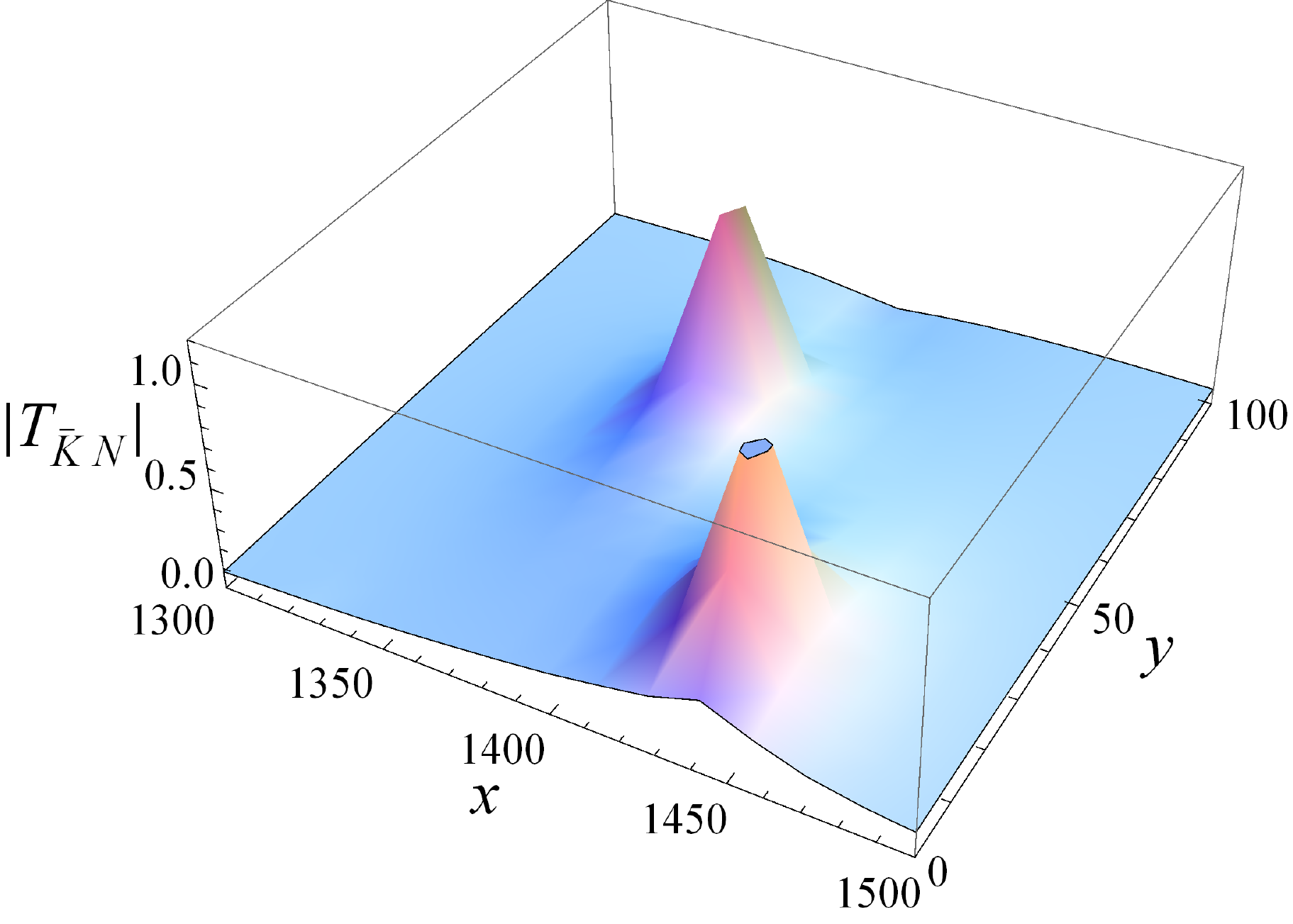} 
	\caption{Modulus of the scattering amplitude $\left|T_{\bar{K} N}\right|$ in the second Riemann sheet of the complex energy plane $(\sqrt{s}=x+i y)$ for the interactions of four (left) and two (right) coupled channels.}
	\label{fig:absTKN}
\end{figure}

In the presence of bound state or resonance, a peak will appear in modulus square of the scattering amplitude. Thus, to look for the resonance of the interaction, we plot the modulus squares of the scattering amplitudes of $\bar{K}N$ and $\pi\Sigma$ for the case of the coupled channel interactions, which are shown in Fig.~\ref{fig:42T}. As one can see from Fig.~\ref{fig:42T}, there are clear peak structures above the $\pi\Sigma$ threshold in all the scattering amplitudes, and the line shape of the curves for two and four coupled channels are consistent with each other, where only the maximum values and the peak positions are a little different in the $\bar{K}N\rightarrow\bar{K}N$ and $\bar{K}N\rightarrow\pi\Sigma$ channels. These results also show that the other two coupled channels $\eta\Lambda$ and $K\Xi$ have small contribution to these peaks. But, the properties of these peaks are different. The one appearing in the $\bar{K} N \rightarrow \bar{K} N$ interaction has a mass close to the $\bar{K}N$ threshold with a narrower width, which is similar to the one of the $\bar{K} N \rightarrow \pi \Sigma$ transition. However, the one from the $\pi \Sigma \rightarrow \pi \Sigma$ interaction has a lower mass with a wider width.  These two poles's coherence behaviour are similar the ones as shown in Ref.~\cite{Sombillo:2021rxv}. Indeed, this is the two-pole structures, as found in Refs.~\cite{Oller:2000fj,Jido:2002yz,GarciaRecio:2002td,Jido:2003cb}, as shown clearly in Fig.~\ref{fig:absTKN}, where we plot the modulus of the scattering amplitude $\left|T_{\bar{K} N}\right|$ in the second Riemann sheet for the two cases of coupled channel interactions. 

In order to get the masses and widths of these peaks, we look for the pole in the unphysical Riemann sheet, which correspond to the singularity in the scattering amplitude in complex energy plane. As discussed in the last section, in the ChUA the Riemann sheets can be defined through the loop function $G_{l}(s)$. By doing that, we find the corresponding poles of these peaks in Fig.~\ref{fig:42T} in the second Riemann sheet, as summarized in Table~\ref{tab:positionofpoles}, where two regularization methods were used for the loop function $G_{l}(s)$, the CO method and DR method. In the case of four coupled channels, using DR method two poles are found in the second Riemann sheet, which correspond to the $\Lambda(1405)$ state 
\footnote{In the latest PDG~\cite{Zyla:2020zbs}, the lower-mass pole was assigned as a $\Lambda(1380)$ resonance and the higher one considered for the $\Lambda(1405)$ state.}. 
Both poles are below the $\bar{K}N$ threshold and above the $\pi\Sigma$ threshold, and thus the $\pi\Sigma$ channel is the only open channel in $I=0$. Another pole is found, which corresponds to the $\Lambda(1670)$ state. These results are consistent with the ones obtained in Ref.~\cite{Jido:2003cb}. However, using CO method two similar poles are found, whereas the third pole is far above the region where the $\Lambda(1670)$ state appeared. The inconsistency for the third pole indicates that using one CO can not reproduced well for both $\Lambda(1405)$ and $\Lambda(1670)$ state, and different CO should be used for the $K\Xi$ interaction to dynamically generate the $\Lambda(1670)$ state. Since we mainly concern about the $\Lambda(1405)$ state, we will use the DR method for the following calculations. Thus, to investigate more details of these two poles for the $\Lambda(1405)$ state, we omit the other two channels $\eta\Lambda$ and $K\Xi$, and study the coupled channel interactions of two channels $\bar{K}N$ and $\pi\Sigma$. Two poles $\sqrt{s_{p}}=(1384.54+60.98i)$ MeV and $\sqrt{s_{p}}=(1438.95+12.44i)$ MeV are present in the second Riemann sheet, which are consistent with the ones obtained by the CO method. Compared with the results of four coupled channels, the mass and width of the lower-mass pole is slightly smaller. On the contrary, for the higher-mass pole the mass is become larger and the width is smaller. From these results, one can see that the position of the poles do not change remarkably by reducing the coupled channels from four channels to two channels. Hence, these two poles are coming from the strong interactions of the $\bar{K}N$ and $\pi\Sigma$ channels, which can also be confirmed by the results of the couplings and the single channel interaction below. Note that the mass of higher-mass pole in two coupled channels case is slightly above the threshold of $\bar{K}N$ channel.

\begin{table}
     \renewcommand{\arraystretch}{1.3}
     \setlength{\tabcolsep}{0.2cm}
	\centering
	\caption{Poles positions
	\footnote{Note that the poles are always a pair of conjugated solutions in the complex Riemann sheet. But, the pole corresponding to a resonance should be the one having negative imaginary part, $\sqrt{s_{p}}=M_{r} - i \Gamma_{r}/2$ as defined above. With the definition of Eq.~\eqref{eq:GII}, one can find the positive imaginary one without strictly specified the signs of the three momenta for the definition of the Riemann sheets, which is sometimes assigned as the pole of the resonance.}
	compared with other works and experimental results (Unit: MeV).}
	\resizebox{1\textwidth}{!}
	{\begin{tabular}{lccc}			
			\hline\hline   &  $1^{st}$ pole   &  $2^{st}$ pole   &  $3^{st}$ pole \\
			\hline 
			Four coupled channels (DR)   &   $1390.61+65.91i$   &   $1426.77+16.32i$   &   $1680.38+20.18i$   \\
			Four coupled channels (CO)   &   $1377.35+58.46i$   &   $1427.96+18.10i$   &   $1762.8 + 27.51 i$  \\
			Two coupled channels (DR)    &   $1384.54+60.98i$   &   $1438.95+12.44i$   &  \\
			Two coupled channels (CO)    &   $1375.56+62.63i$   &   $1438.04+11.44i$   & \\
			Ref.~\cite{Ikeda:2012au}, NLO         &   $1381_{-6}^{+18}-81_{-8}^{+19} i$   &   $1424_{-23}^{+7}-26_{-14}^{+3} i$   &  \\
			Ref.~\cite{Guo:2012vv} solution II   &   $1388_{-9}^{+9}-114_{-25}^{+24} i$   &   $1421_{-2}^{+3}-19_{-5}^{+8} i$   &  \\
			Ref.~\cite{Mai:2014xna} solution II   &   $1330_{-5}^{+4}-56_{-11}^{+17} i$   &   $1434_{-2}^{+2}-10_{-1}^{+2} i$   &  \\
			Ref.~\cite{Mai:2014xna} solution IV   &   $1325_{-15}^{+15}-90_{-18}^{+12} i$   &   $1429_{-7}^{+8}-12_{-3}^{+2} i$   &  \\
			Ref.~\cite{Anisovich:2020lec} one-pole solution   &   &   $(1421 \pm 3)-(23 \pm 3) i$   &  \\
			Ref.~\cite{Anisovich:2020lec} two-pole solution   &   $1380-90 i$ (fixed)   &   $(1423 \pm 3)-(20 \pm 3) i$   &  \\
			\hline\hline			
	\end{tabular}}	
	\label{tab:positionofpoles}
\end{table}

From the former results, we can estimate that each pole mainly contributed from which channels. To show how the poles couple to the channels quantitatively, we calculated the couplings. The coupling strength of the resonance to each coupled channel is extracted from the residues of the scattering amplitude around the pole position as discussed in last section. The corresponding couplings for four coupled channels and two coupled channels are present in Table~\ref{tab:couplings}. For the case of four coupled channel interactions, the couplings of the two channels $\bar{K}N$ and $\pi\Sigma$ for the lower-mass pole $\sqrt{s_{p}}=1390.61+65.91i$ are about three times larger than those of the other two coupled channels, $\eta \Lambda$ and $K \Xi$. Among them, the one of the $\pi\Sigma$ channel is bigger than that of the $\bar{K}N$ channel, which means that the lower-mass pole is dominant by the $\pi\Sigma$  channel. The coupling of the $\bar{K}N$ channel for the higher-mass pole $\sqrt{s_{p}}=1426.77+16.32i$ is about 2.71, which is two times bigger than the ones for the channels $\pi\Sigma$ and $\eta\Lambda$, whereas the one of the last $K\Xi$ channel is quit small. Thus, the higher-mass pole is mainly contributed by the $\bar{K}N$ channel. Besides, the third pole, which corresponds to the $\Lambda(1670)$ state, is strongly coupled to the $K \Xi$ channel. These results are compatible with the results of Refs.~\cite{Jido:2003cb,Sekihara:2010uz} and crossed check by two coupled channels results.

Using the sum rule of Eq.~\eqref{eq:components}, the compositeness of the two poles was computed with the couplings obtained above, where one can check whether these poles are a pure molecular state or have something else. Our results are shown in Table~\ref{tab:compositeness}. As one can see in Table~\ref{tab:compositeness}, the first pole is highly dominated by the $\pi\Sigma$ components with the amount of 0.63, whereas the one of $\bar{K}N$ is about three times smaller. The components of second pole are mainly constructed by the part of $\bar{K}N$, which is up to 98\%, and have a tiny part of $\pi\Sigma$. Note that the contributions from the other two channels, $\eta \Lambda$ and $K \Xi$, are quite small for these two poles, owing to their thresholds far away above two poles. Indeed, this conclusion can be confirmed by the results of the two coupled channels. Besides, in the results of four coupled channels, there is the third pole, corresponding to the $\Lambda(1670)$ state, which is significantly contributed by the $K \Xi$ channel. Note that, as discussed in Ref.~\cite{Matuschek:2020gqe}, the compositeness condition of Eq.~\eqref{eq:components} is valid for the case of the weak-binding limit and the pole with narrow width. Thus, the compositeness of the third pole is questionable, since it is the bound state of the $K \Xi$ with large binding energy of about 134 MeV. Indeed, this is similar to the case of the $f_0(500)$ (or called $\sigma$) resonance as found in Ref.~\cite{Ahmed:2020kmp}, which has a large width of about 400 MeV. This is why we investigate the couplings and the compositeness together. Thus, our final conclusions are based on both results.

\begin{table}
 \renewcommand{\arraystretch}{1.3}
	\centering
	\caption{Couplings for each coupled channel compared with the other works.}
	\resizebox{1\textwidth}{!}
	{\begin{tabular}{ccccccccc}
			\hline \hline 
			$\sqrt{s_{p}}$ & $g_{\bar{K} N}$ & $\left|g_{\bar{K}N}\right|$ & $g_{\pi \Sigma}$ & $\left|g_{\pi \Sigma}\right|$ & $g_{\eta \Lambda}$ &$\left|g_{\eta \Lambda}\right|$ & $g_{K \Xi}$& $\left|g_{K \Xi}\right|$ \\
			\hline 
			\makecell[c]{Four coupled channels} \\
			$1390.61+65.91i$ & $1.20+1.74i$ & $2.12$ & $2.46+1.51i$ & $2.88$ & $0.01+0.76i$ & $0.76$ & $0.45+0.41i$ & $0.61$ \\
			$1426.77+16.32i$ & $2.55-0.94i$ & $2.71$ & $0.42-1.45i$ & $1.51$ & $1.40-0.20i$ & $1.41$ & $0.11-0.33i$ & $0.35$ \\
			$1680.38+20.18i$ & $0.30+0.72i$ & $0.78$ & $0.01+0.27i$ & $0.27$ & $1.05+0.11i$ & $1.06$ & $3.45+0.12i$ & $3.46$ \\
			\makecell[c]{Two coupled channels} \\
			$1384.54+60.98i$ & $1.46+1.45i$ & $2.06$ & $2.38+1.29i$ & $2.71$ & $-$ & $-$ & $-$ & $-$ \\
			$1438.95+12.44i$ & $2.00-1.34i$ & $2.41$ & $0.07+1.42i$ & $1.42$ & $-$ & $-$ & $-$ & $-$ \\
			\makecell[c]{Ref.~\cite{Jido:2003cb}} \\
			$1390+66i$ & $1.2+1.7i$ & $2.1$ & $-2.5-1.5i$ & $2.9$ & $0.010+0.77i$ & $0.77$ & $-0.45-0.41i$ & $0.61$ \\
			$1426+16i$ & $-2.5+0.94i$ & $2.7$ & $0.42-1.4i$ &$1.5$ & $-1.4+0.21i$ & $1.4$ & $0.11-0.33i$ & $0.35$ \\
			$1680+20i$ & $0.30+0.71i$ & $0.77$ & $-0.003-0.27i$ & $0.27$ & $-1.1-0.12i$ & $1.1$ & $3.4+0.14i$ & $3.5$ \\			
			\makecell[c]{Ref.~\cite{Sekihara:2010uz}} \\
			$1391-66i$ & $-0.86+1.26i$ & $-$ & $-1.42+0.88i$ & $-$ & $-0.01+0.79i$ & $-$ & $-0.33+0.30i$ & $-$ \\
			$1426-17i$ & $1.84+0.67i$ & $-$ & $0.26+0.85i$ &$-$ & $1.44+0.21i$ & $-$ & $0.09+0.24i$ & $-$ \\
			\hline\hline    
	\end{tabular}}
	\label{tab:couplings}
\end{table}

\begin{table}
 \renewcommand{\arraystretch}{1.3}
	\centering
	\caption{Compositeness of the poles in coupled channels approach.}
	\resizebox{1\textwidth}{!}
	{\begin{tabular}{ccccccccc}
			\hline \hline 
			$\sqrt{s_{p}}$ & $(1-Z)_{\bar{K} N}$ & $\left|(1-Z)_{\bar{K}N}\right|$ & $(1-Z)_{\pi \Sigma}$ & $\left|(1-Z)_{\pi \Sigma}\right|$ & $(1-Z)_{\eta \Lambda}$ &$\left|(1-Z)_{\eta \Lambda}\right|$ & $(1-Z)_{K \Xi}$& $\left|(1-Z)_{K \Xi}\right|$ \\
			\hline 
			\makecell[c]{Four coupled channels} \\
			$1390.61+65.91i$ & $-0.20+0.12i$ & $0.24$ & $0.36-0.52i$ & $0.63$ & $-0.01-0.002i$ & $0.01$ & $-0.0003+0.01i$ & $0.01$ \\
			$1426.77+16.32i$ & $0.98-0.04i$ & $0.98$ & $-0.04+0.15i$ & $0.15$ & $0.05-0.01i$ & $0.05$ & $-0.002-0.001i$ & $0.002$ \\
			$1680.38+20.18i$ & $0.02+0.002i$ & $0.02$ & $0.002+0.002i$ & $0.003$ & $-0.07-0.16i$ & $0.17$ & $0.52+0.09i$ & $0.52$ \\
			\makecell[c]{Two coupled channels} \\
			$1384.54+60.98i$ & $-0.12+0.18i$ & $0.22$ & $0.27-0.51i$ & $0.58$ & $-$ & $-$ & $-$ & $-$ \\
			$1438.95+12.44i$ & $0.96-0.09i$ & $0.97$ & $0.05+0.12i$ & $0.13$ & $-$ & $-$ & $-$ & $-$ \\
			\hline\hline    
	\end{tabular}}
	\label{tab:compositeness}
\end{table}

\begin{figure}
		\centering
		\includegraphics[width=0.49\linewidth]{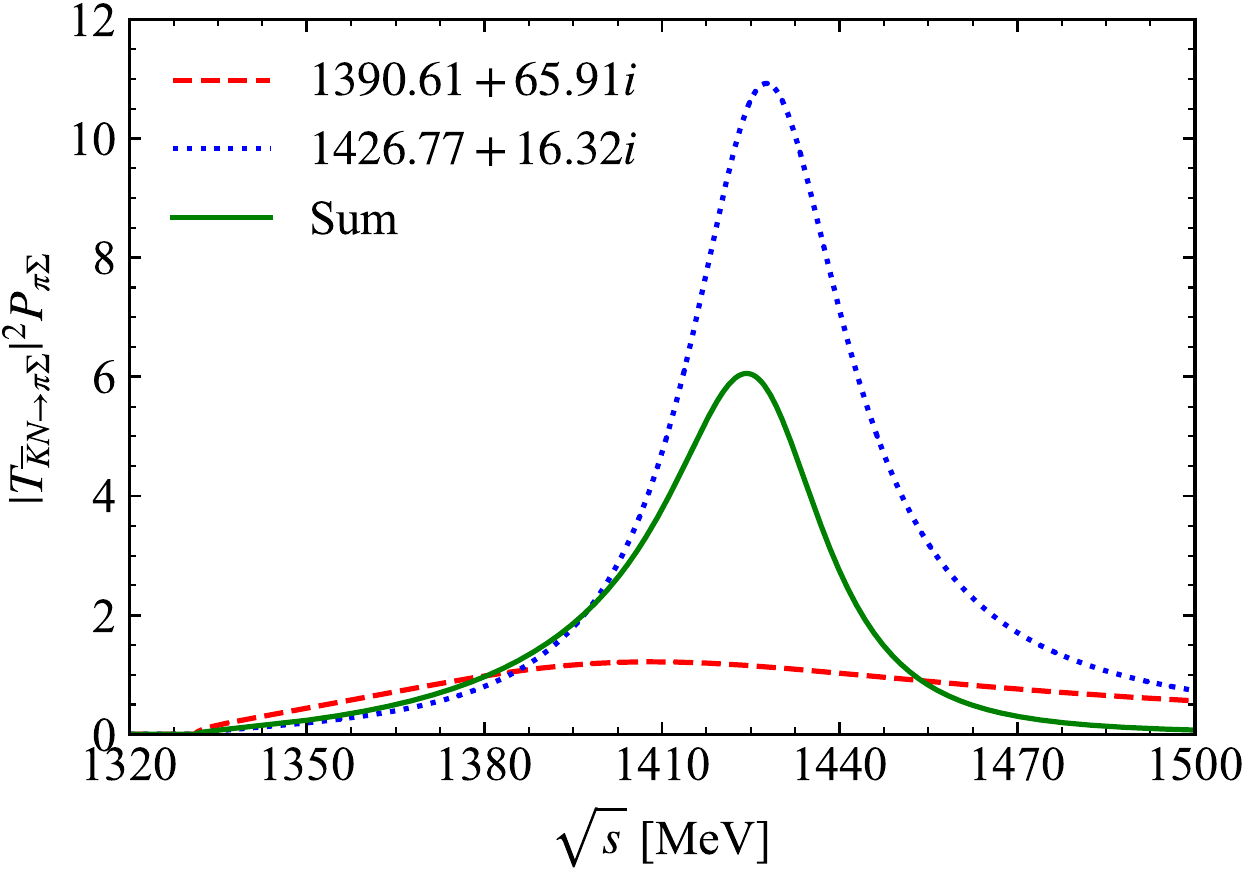} 
		\includegraphics[width=0.49\linewidth]{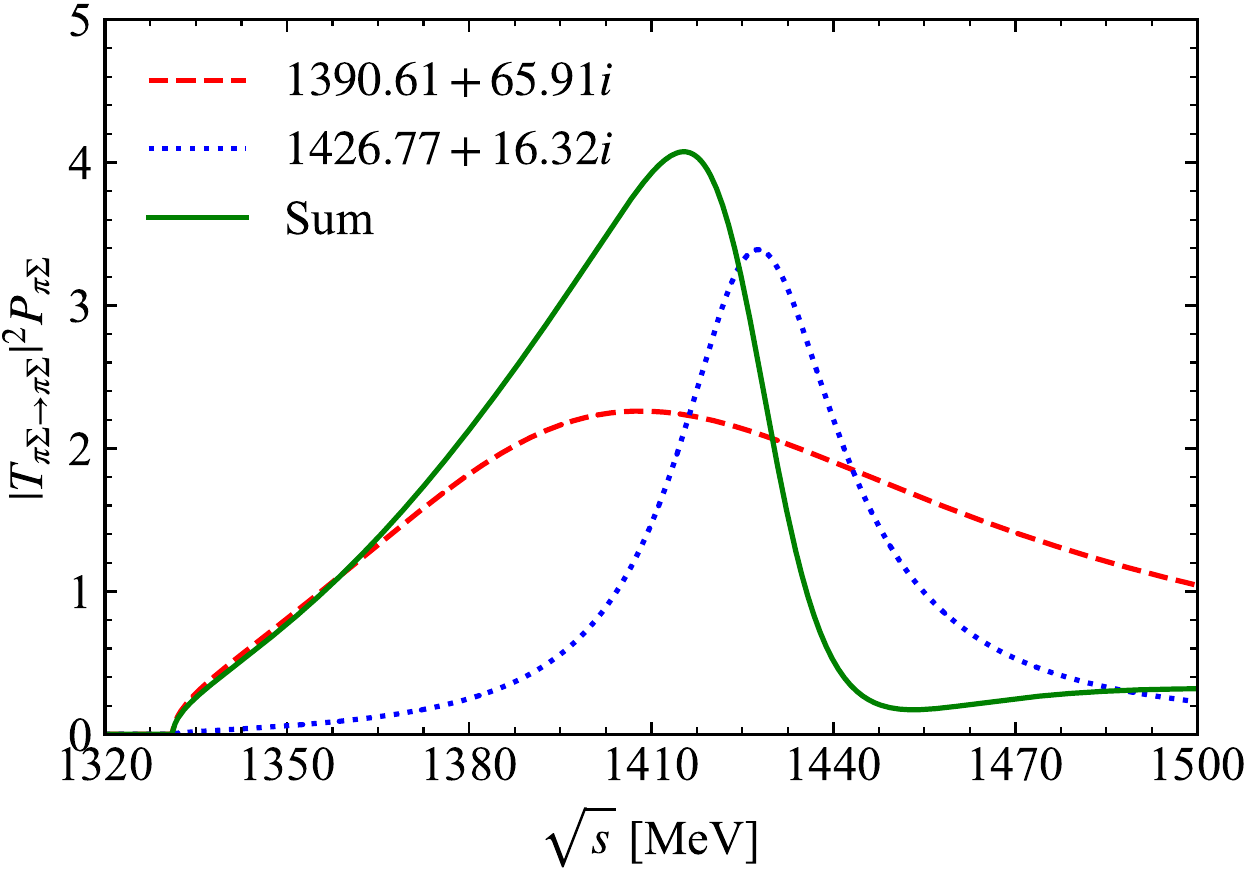} 
		\includegraphics[width=0.49\linewidth]{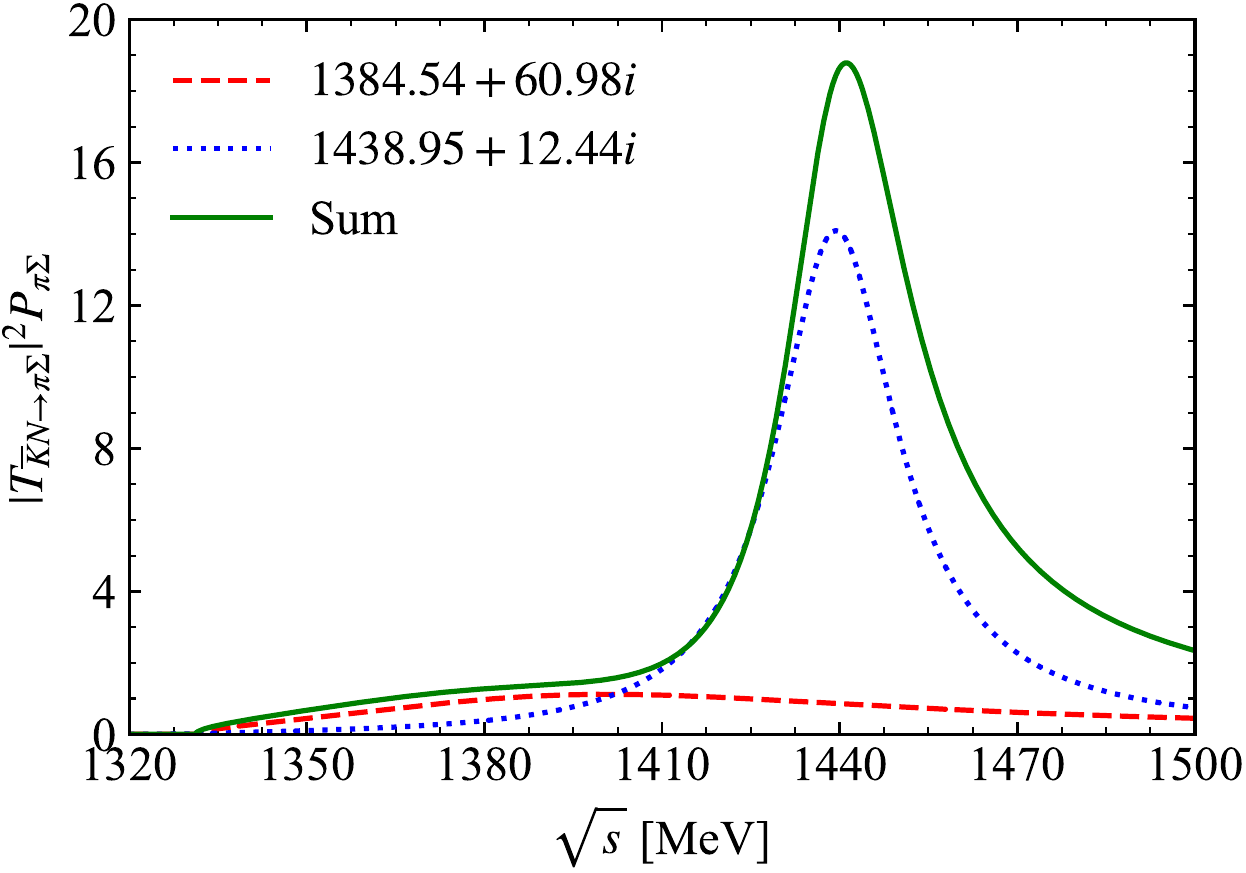}  
		\includegraphics[width=0.49\linewidth]{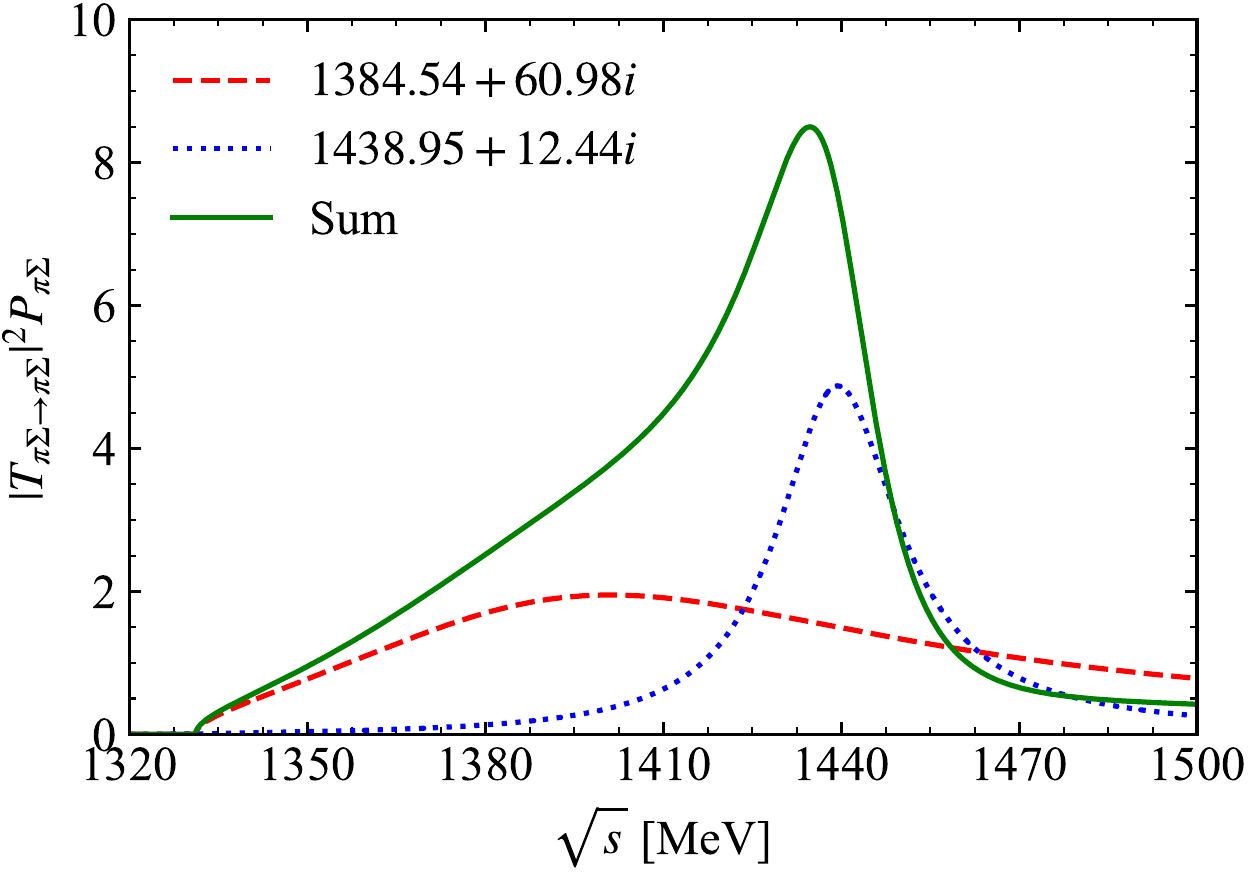}  
	\caption{The $\pi\Sigma$ spectra using Breit-Wigner parameterization with Eqs.~\eqref{eq:Breit-Wigner11} and \eqref{eq:Breit-Wigner12} for four coupled channels (upper) and two coupled channels (lower).}
	\label{fig:42AF}
\end{figure}

\begin{figure}
		\centering
		\includegraphics[width=0.49\linewidth]{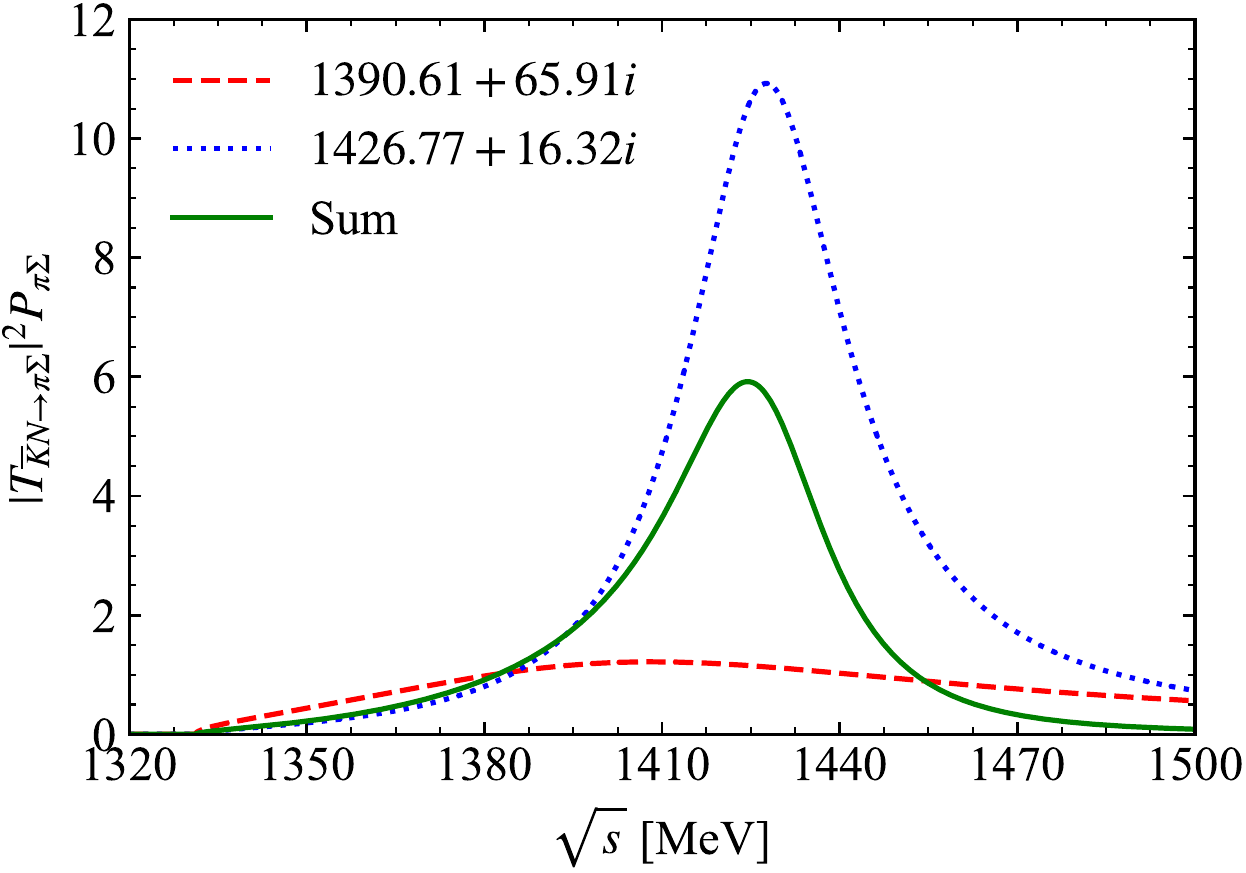} 
		\includegraphics[width=0.49\linewidth]{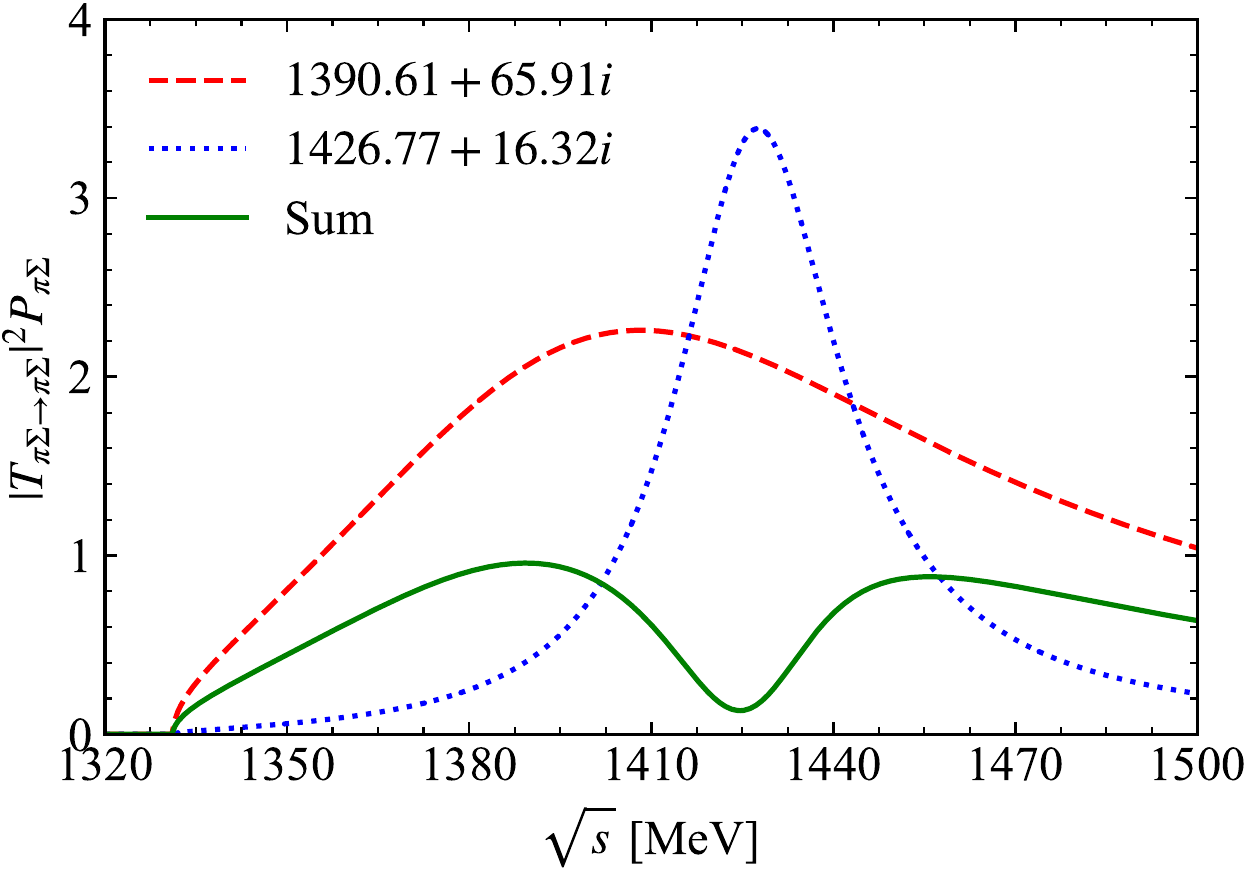} 
		\includegraphics[width=0.49\linewidth]{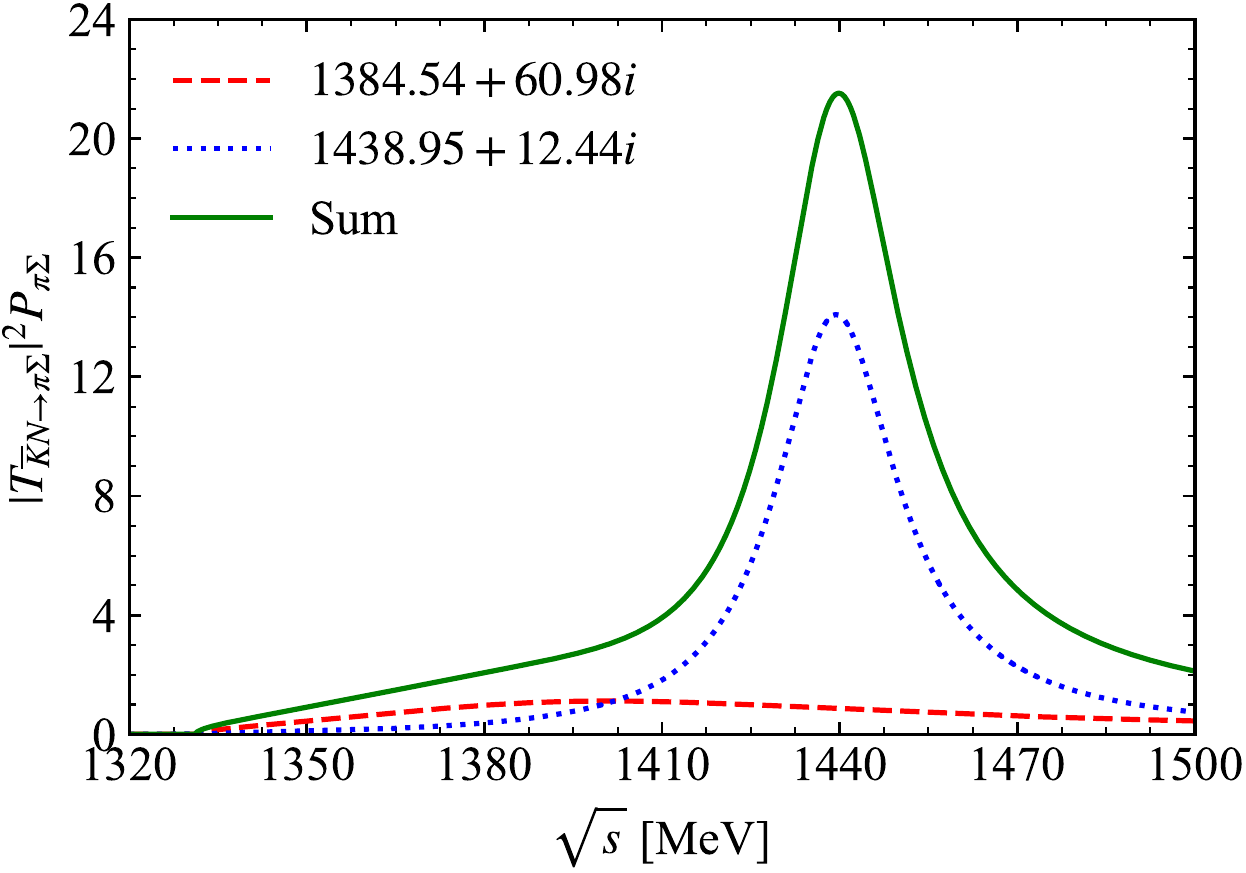}  
		\includegraphics[width=0.49\linewidth]{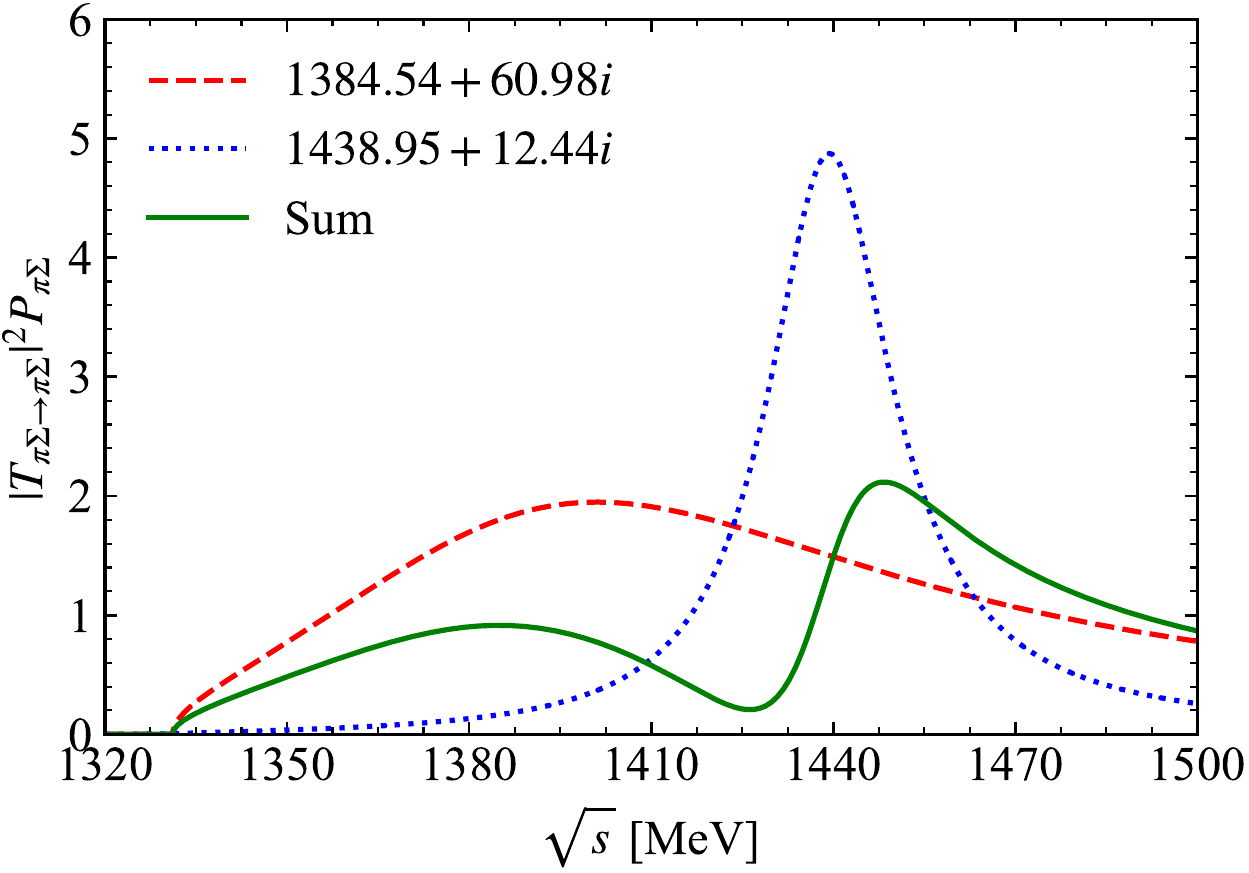}  
	\caption{The same as Fig.~\ref{fig:42AF} but with Eqs.~\eqref{eq:Breit-Wigner21} and \eqref{eq:Breit-Wigner22} for four coupled channels (upper) and two coupled channels (lower).}
	\label{fig:42AFm}
\end{figure}

The CLAS collaboration reported the observation of two pole structure of $\Lambda(1405)$ from the electroproduction process, and found that this state was not a simple Breit-Wigner resonance~\cite{Lu:2013nza}. One can see from the results of Fig.~\ref{fig:42T} that the scattering amplitudes of $T_{\pi \Sigma \to \pi \Sigma}$ and $T_{\bar{K} N \to \bar{K} N}$ ($T_{\bar{K} N \to \pi \Sigma}$) are dominated by the lower and higher-mass poles, respectively. As discussed in Ref.~\cite{Hyodo:2011ur}, as a consequence of the interference of these two poles in the amplitudes, one could observe only a single resonance peak experimentally in one channel. Therefore, if the $\Lambda(1405)$ is not a single resonance, it should be a superposition of two independent states with the same quantum numbers. To show this interference effect on the final observation, as done in Ref.~\cite{Jido:2003cb}, we take
\begin{equation}
	T_{\bar{K} N \rightarrow \pi \Sigma}=g_{\bar{K} N}^{p_{1}} \frac{1}{\sqrt{s}-M_{p_{1}}+i \Gamma_{p_{1}} / 2} g_{\pi \Sigma}^{p_{1}}+g_{\bar{K} N}^{p_{2}} \frac{1}{\sqrt{s}-M_{p_{2}}+i \Gamma_{p_{2}} / 2} g_{\pi \Sigma}^{p_{2}},
	\label{eq:Breit-Wigner11}
\end{equation}
\begin{equation}
	T_{\pi \Sigma \rightarrow \pi \Sigma}=g_{\pi \Sigma}^{p_{1}} \frac{1}{\sqrt{s}-M_{p_{1}}+i \Gamma_{p_{1}} / 2} g_{\pi \Sigma}^{p_{1}}+g_{\pi \Sigma}^{p_{2}} \frac{1}{\sqrt{s}-M_{p_{2}}+i \Gamma_{p_{2}} / 2} g_{\pi \Sigma}^{p_{2}},
	\label{eq:Breit-Wigner12}
\end{equation}
where the upper indexes $p_1$ and $p_2$ denote the first pole and second pole found in the second Riemann sheet, respectively, and $g_i^p$ are the couplings of $i$-channel for a certain pole ($p$), which were given in Table~\ref{tab:couplings}. The results are shown in Fig.~\ref{fig:42AF}, where one can see that the results of Eq.~\eqref{eq:Breit-Wigner12} for four coupled channels (the top-right one) are well consistent with the one of $\pi\Sigma$ invariant mass distributions in Fig.~\ref{fig:42Gamma}. For two coupled channels, the higher mass pole looks like to have stronger strengths. Indeed, the results of Eq.~\eqref{eq:Breit-Wigner11} are mainly contributed by the higher-mass pole, which are consistent with the ones of Fig.~\ref{fig:42T} for both four channels and two channels cases. Furthermore, to show the inference effect more clearly, we take for the conjugated pole,
\begin{equation}
	T_{\bar{K} N \rightarrow \pi \Sigma}=g_{\bar{K} N}^{p_{1}} \frac{1}{\sqrt{s}-M_{p_{1}}-i \Gamma_{p_{1}} / 2} g_{\pi \Sigma}^{p_{1}}+g_{\bar{K} N}^{p_{2}} \frac{1}{\sqrt{s}-M_{p_{2}}-i \Gamma_{p_{2}} / 2} g_{\pi \Sigma}^{p_{2}},
	\label{eq:Breit-Wigner21}
\end{equation}
\begin{equation}
	T_{\pi \Sigma \rightarrow \pi \Sigma}=g_{\pi \Sigma}^{p_{1}} \frac{1}{\sqrt{s}-M_{p_{1}}-i \Gamma_{p_{1}} / 2} g_{\pi \Sigma}^{p_{1}}+g_{\pi \Sigma}^{p_{2}} \frac{1}{\sqrt{s}-M_{p_{2}}-i \Gamma_{p_{2}} / 2} g_{\pi \Sigma}^{p_{2}},
	\label{eq:Breit-Wigner22}
\end{equation}
where the results are shown in Fig.~\ref{fig:42AFm}. From Figs.~\ref{fig:42AF} and \ref{fig:42AFm}, one can see that the peak position of the $\Lambda(1405)$ is mainly contributed by the two poles, due to their interference consequence, which is originated from the strong coupled effect between the channels of $\bar{K} N$ and $\pi \Sigma$. Thus, the mass and width of the peak $\Lambda(1405)$ are between two poles' masses and widths, respectively. For more discussions about the interference effects of the Breit-Wigner parameterization can be referred to Ref.~\cite{Bai:2019jrb}.

\begin{figure}
\centering 
		\includegraphics[width=0.49\linewidth]{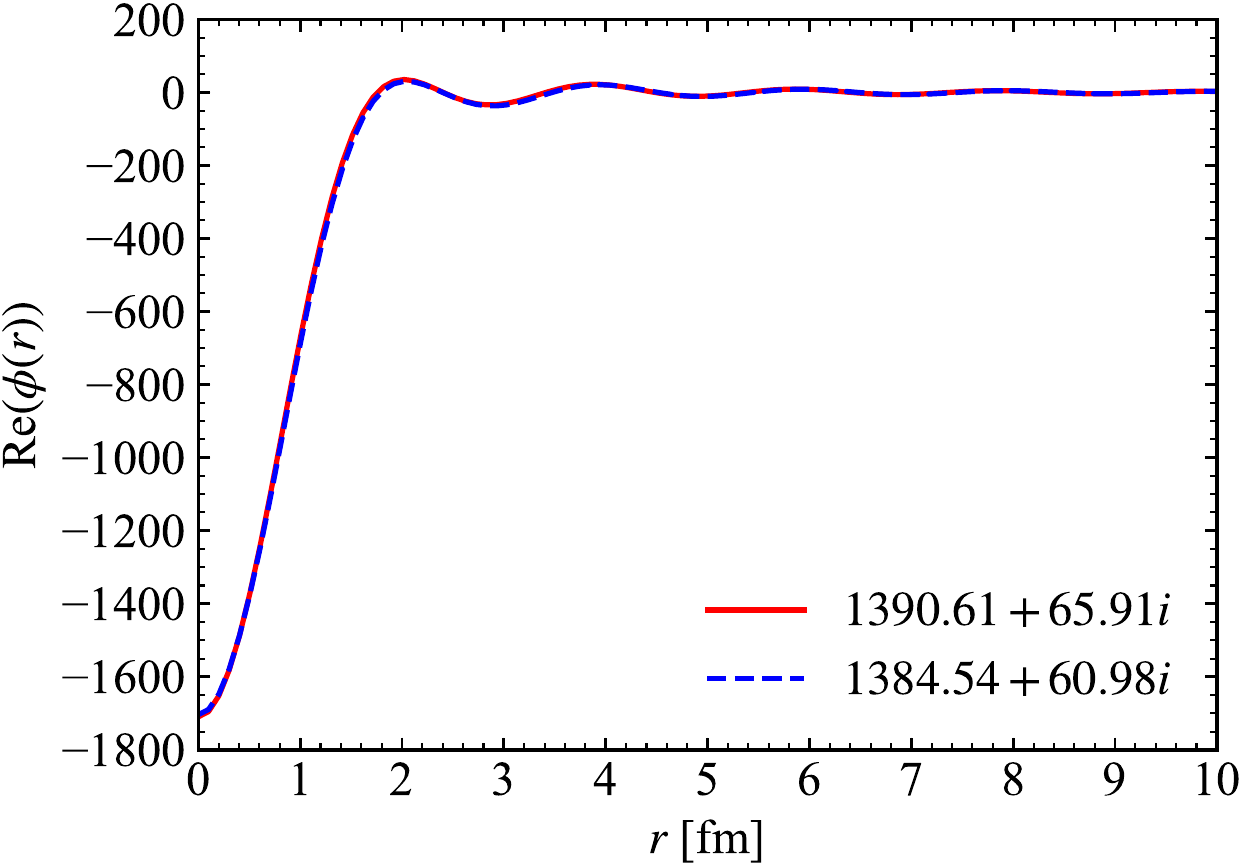} 
		\includegraphics[width=0.49\linewidth]{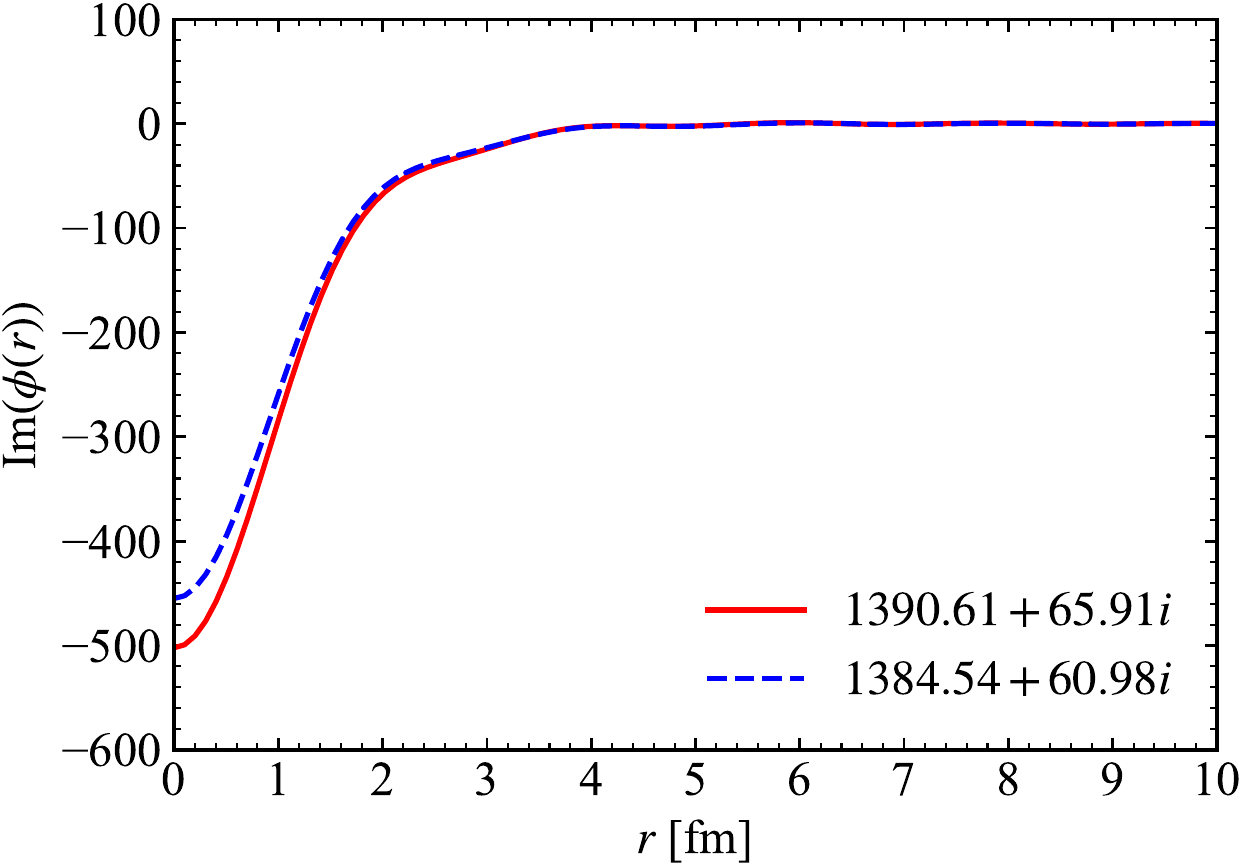} 
		\includegraphics[width=0.49\linewidth]{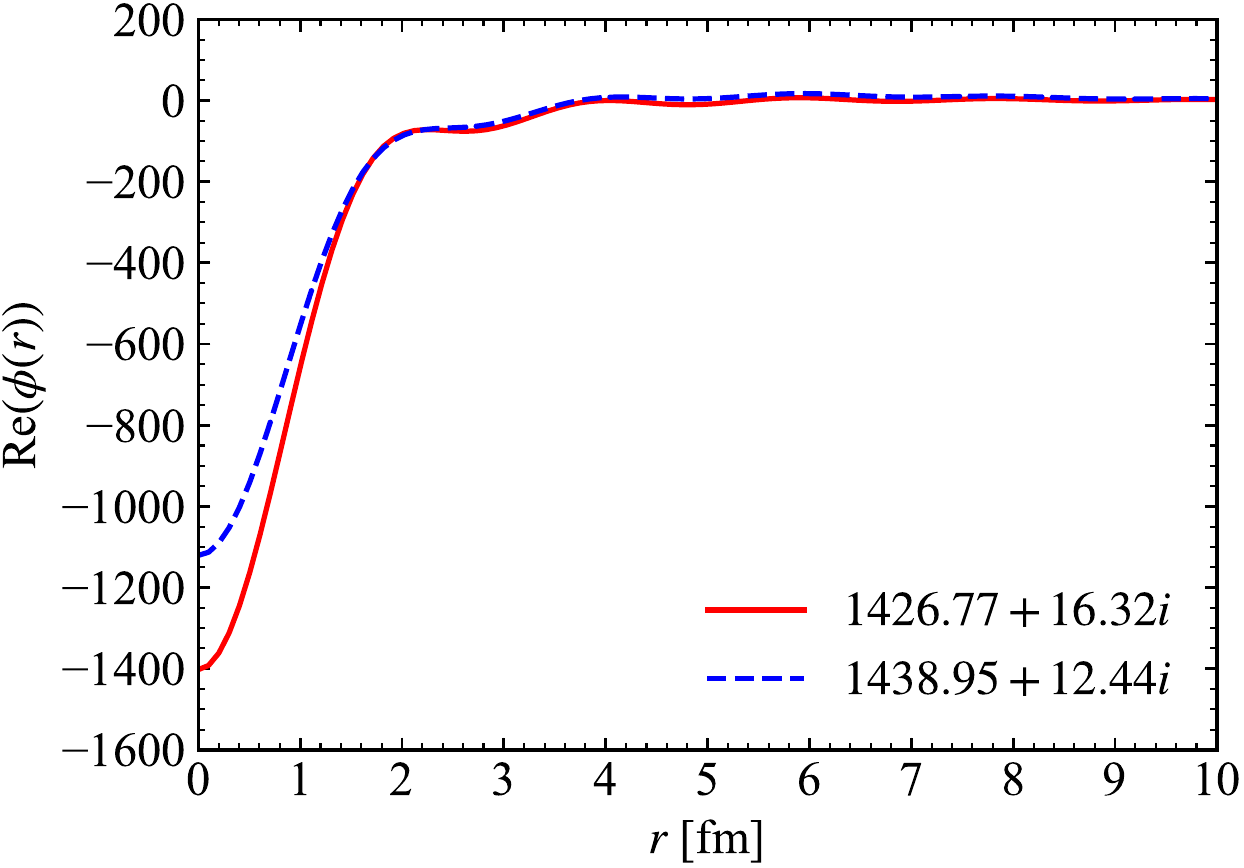} 
		\includegraphics[width=0.49\linewidth]{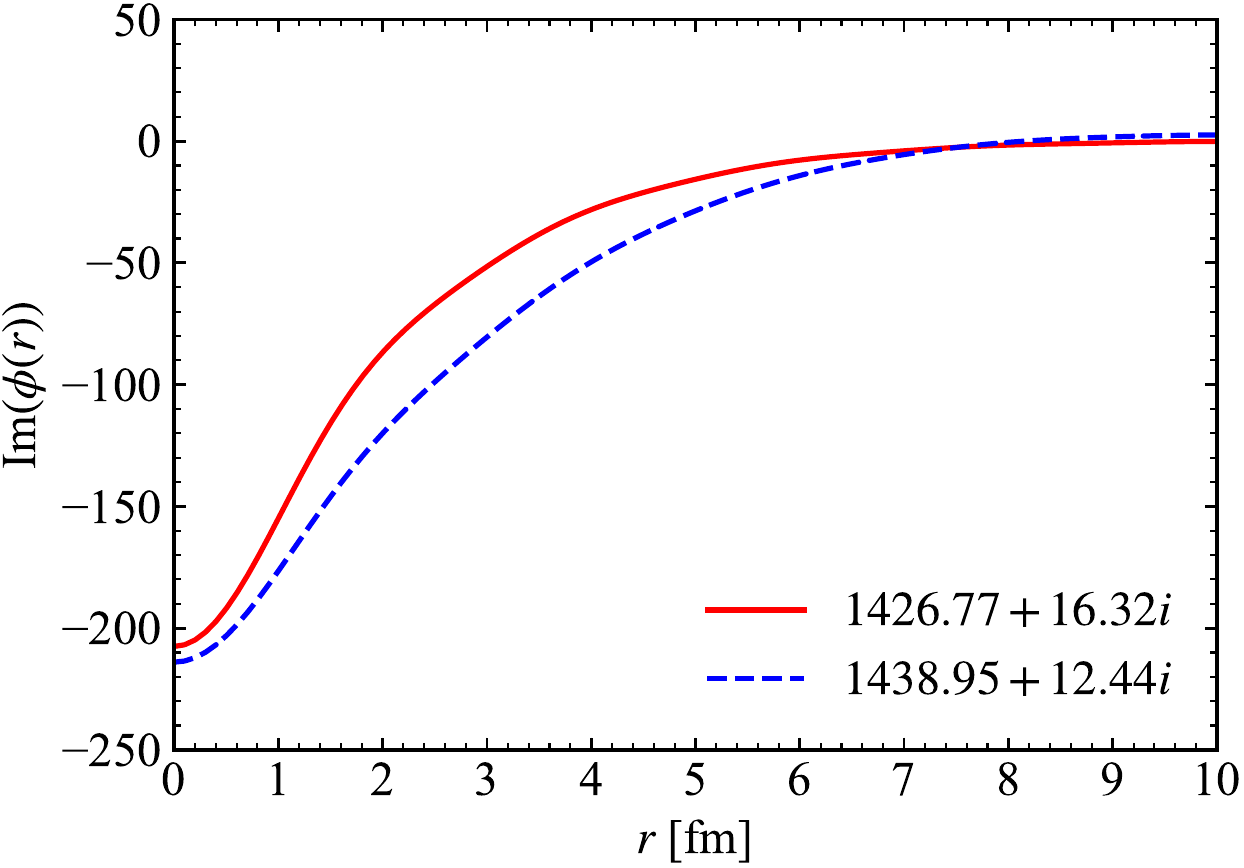} 
	\caption{Wave functions of two poles for the $\bar{K}N$ channel.}
	\label{fig:42WFKN}
\end{figure}

\begin{figure}
		\centering
		\includegraphics[width=0.49\linewidth]{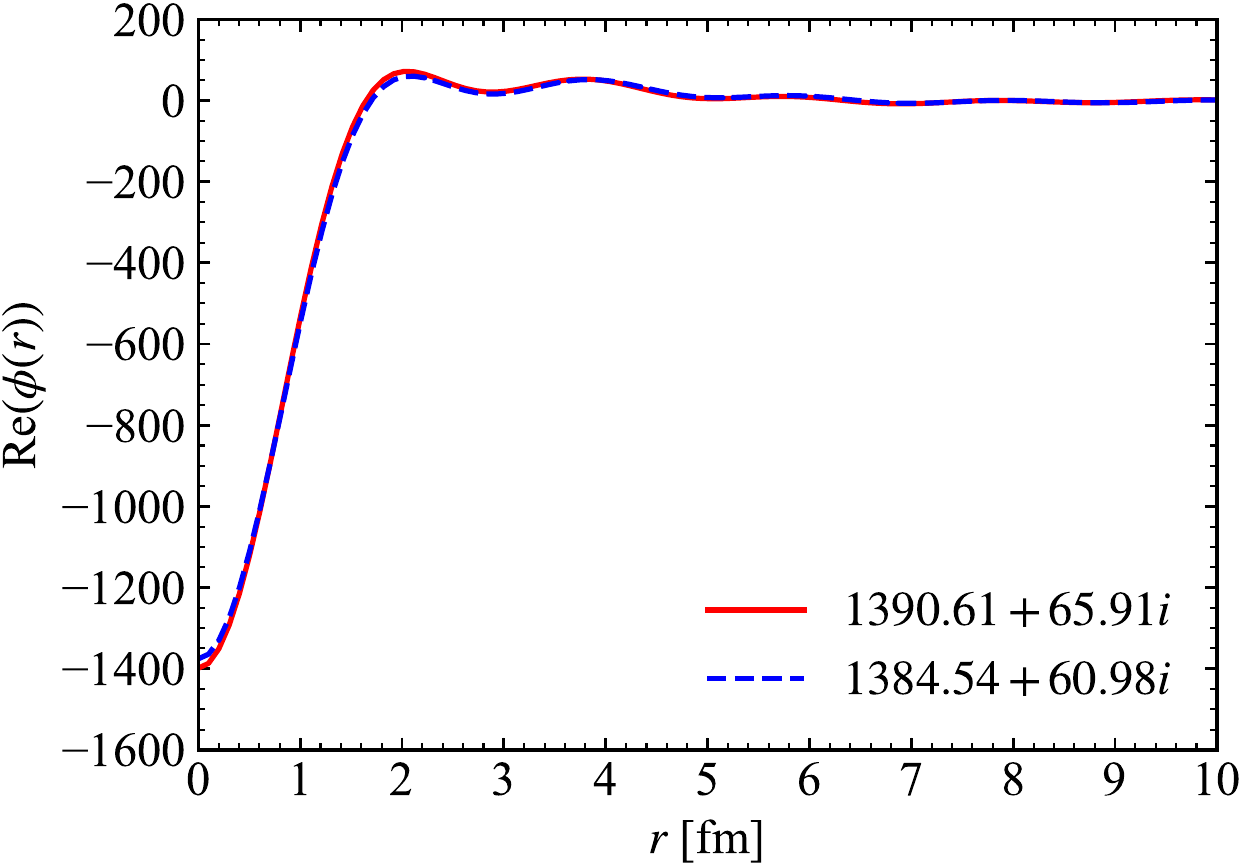} 
		\includegraphics[width=0.49\linewidth]{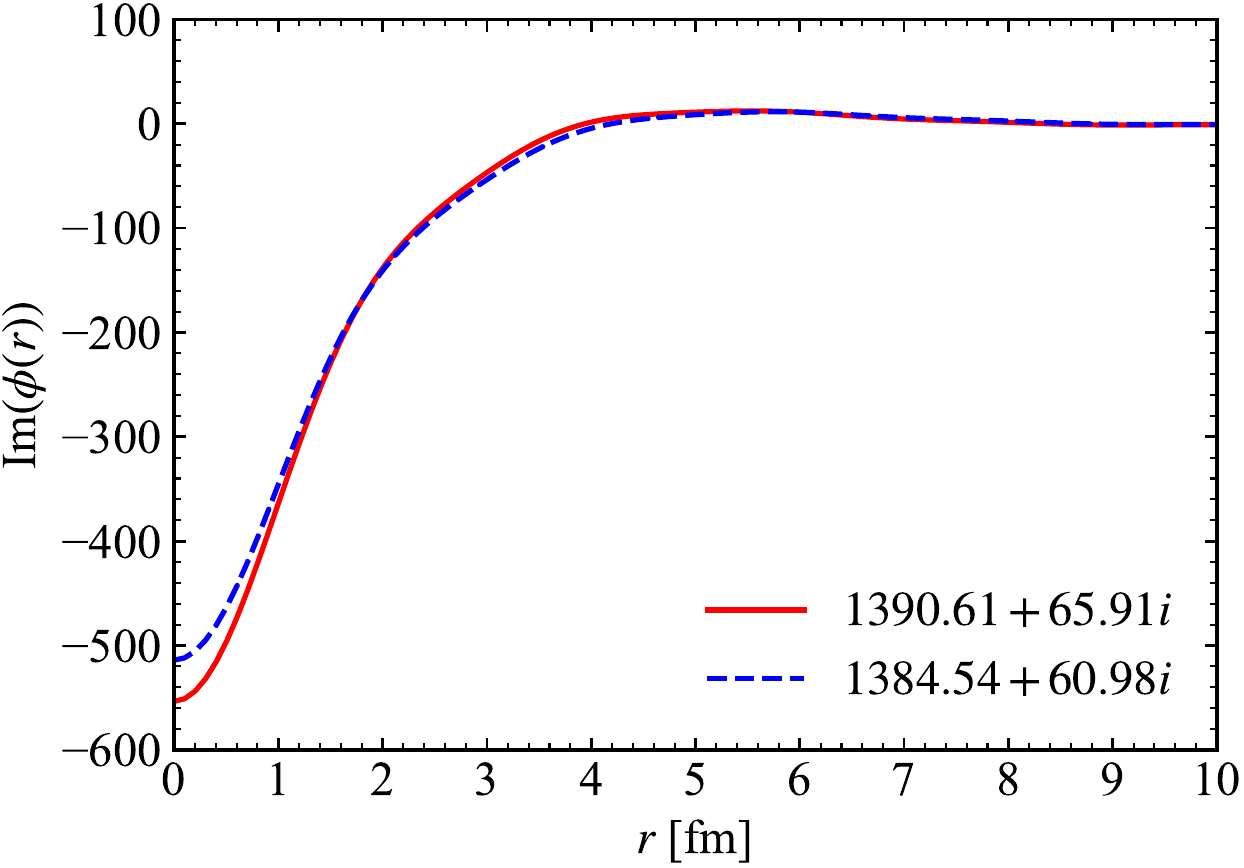} 
		\includegraphics[width=0.49\linewidth]{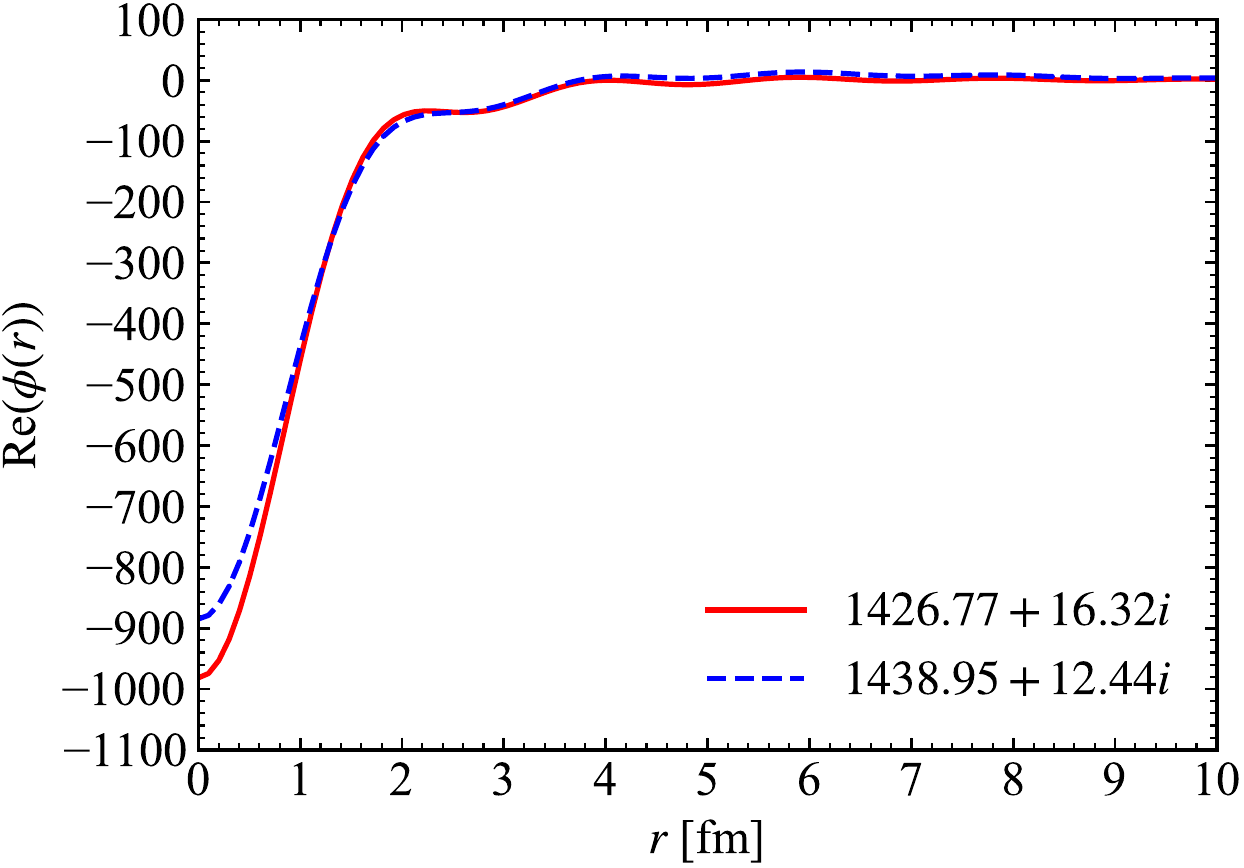} 
		\includegraphics[width=0.49\linewidth]{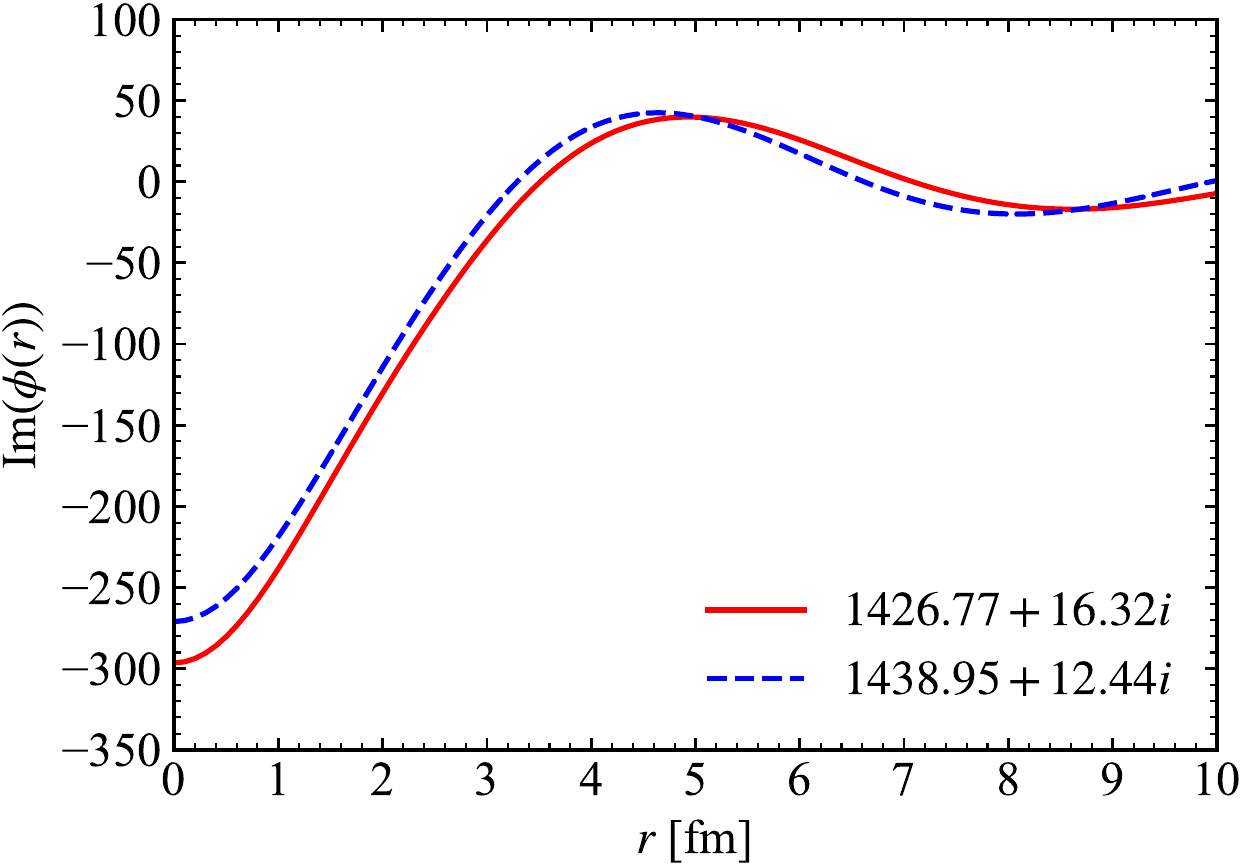} 
	\caption{Wave functions of two poles for the $\pi\Sigma$ channel.}
	\label{fig:42WFPS}
\end{figure}

The wave function also contains the information of the structure of a state, since the wave function in coordinate space indicates the spatial composition of the components. The wave functions of the two poles for the cases of four and two coupled channels are shown in Figs.~\ref{fig:42WFKN} and \ref{fig:42WFPS} for the components of $\bar{K} N$ and $\pi \Sigma$, respectively, which are consistent with the ones obtained in Ref.~\cite{YamagataSekihara:2010pj} qualitatively (with different normalization). From Figs.~\ref{fig:42WFKN} and \ref{fig:42WFPS}, one can see that the real part of wave functions are almost become zero after a certain size of about 4 fm, and the imaginary parts are similar, which means that the wave functions would be confined within a few fm. Note that the results of Figs.~\ref{fig:42WFKN} and \ref{fig:42WFPS} are just the scattering wave functions. In principle, for the case of the open channel, the $\pi \Sigma$ channel in our case, the wave functions in Fig.~\ref{fig:42WFPS} would have the parts of the outgoing plane wave function, which were out of our concerns and ignored. More details can be found in Refs.~\cite{YamagataSekihara:2010pj,Aceti:2012dd}.

To quantify the spatial size of the $\Lambda(1405)$, we calculated the radii of states using Eq.~\eqref{eq:radii} with the form factors obtained. Since the form factors are complex with the poles of the resonance having widths, the radii for two poles become complex, as shown in Table~\ref{tab:radii}. The modulus of radius of the higher-mass pole is $1.57$ fm and $1.58$ fm for four and two coupled channels, respectively. However, the one for the lower-mass pole is $0.67$ fm and $0.66$ fm, respectively, which are within the typical hadronic scale $\leq0.8$ fm~\cite{Sekihara:2012xp}. These results for the higher-mass pole are consistent with the ones obtained in Ref.~\cite{Sekihara:2012xp}, about 1.69 fm with binding energy 10 MeV. Consequently, the radius of the higher-mass pole is about two times larger than the typical size of the hadrons, which has quite different size nature compared with the one for the low-mass pole.

\begin{table}
 \renewcommand{\arraystretch}{1.3}
	\centering
	\caption{Radii of the poles evaluated with Eq.~\eqref{eq:radii} in coupled channels approach.}
	\begin{tabular}{ccccc}
			\hline \hline 
			$\sqrt{s_{p}}$ & $\sqrt{\left\langle r^{2}\right\rangle_{\bar{K}N}}$ & $\mid \sqrt{\left\langle r^{2}\right\rangle_{\bar{K}N} }\mid$ & $\sqrt{\left\langle r^{2}\right\rangle_{\pi \Sigma}}$ & $\mid \sqrt{\left\langle r^{2}\right\rangle_{\pi \Sigma}}\mid$  \\
			\hline 
			\makecell[c]{Four coupled channels} \\
%			\hline 
			$1390.61+65.91i$ & $-$ & $-$ & $(0.67+0.06i)$ fm & $0.67$ fm \\
			$1426.77+16.32i$ & $(1.46+0.57i)$ fm & $1.57$ fm & $-$ & $-$ \\
			\hline
			\makecell[c]{Two coupled channels}  \\
%			\hline 
			$1384.54+60.98i$ & $-$ & $-$ & $(0.65+0.10i)$ fm &$0.66$ fm \\
			$1438.95+12.44i$ & $(1.16+1.07i)$ fm & $1.58$ fm & $-$ & $-$ \\
			\hline\hline    
	\end{tabular}
	\label{tab:radii}
\end{table}

Moreover, we can study the scattering length of the elastic channels at the threshold~\cite{Oset:1997it}, written as
\begin{equation}
a_{i i}=\frac{-1}{4 \pi} \frac{M}{\sqrt{s}} T_{i i},
\label{eq21}
\end{equation}
which is an important parameter in analyzing low energy interactions. The scattering lengths of $K^{-} p$ channel are summarized in Table \ref{scatlen}. The result of four coupled channels is well consistent with the experimental values and the one of next to leading order calculations. Its value is slightly changed for two coupled channels case.

\begin{table}
 \renewcommand{\arraystretch}{1.3}
\center
\caption{\footnotesize Scattering length of $K^{-} p$ channel.} \label{scatlen}
\begin{tabular}{lc}
\hline\hline
       Approaches                            & $a_{K^{-} p} $ (fm)   \\ \hline 
 Four coupled channels                      &  $-0.73 + 1.31 i$       \\ 
 Two coupled channels                       &  $-0.15 + 1.48 i$      \\ 
 Ref.~\cite{Ikeda:2012au}, NLO       &  $-0.70 +  0.89 i$    \\ 
 Ref.~\cite{Guo:2012vv} (Fit I)            &   $-0.67^{+0.13}_{-0.12} +  0.92^{+0.07}_{-0.08} i$   \\
 Ref.~\cite{Guo:2012vv} (Fit II)           &  $-0.61^{+0.13}_{-0.15} +  0.89^{+0.07}_{-0.06} i$  \\
 Exp.~\cite{Iwasaki:1997wf}       &  $ (-0.78 \pm 0.15 \pm 0.03) + (0.49 \pm 0.25 \pm 0.12) i$          \\   
 Exp.~\cite{Martin:1980qe}        &  $ -0.67 + 0.64 i$                            \\   
 Exp~\cite{Ohnishi:2015iaq}         & $ (-0.65 \pm 0.10)+(0.81 \pm 0.15) i  $          \\
 \hline\hline
\end{tabular}
\end{table}

\subsection{Single channel approach}

\begin{figure}
		\centering
		\includegraphics[width=0.49\linewidth]{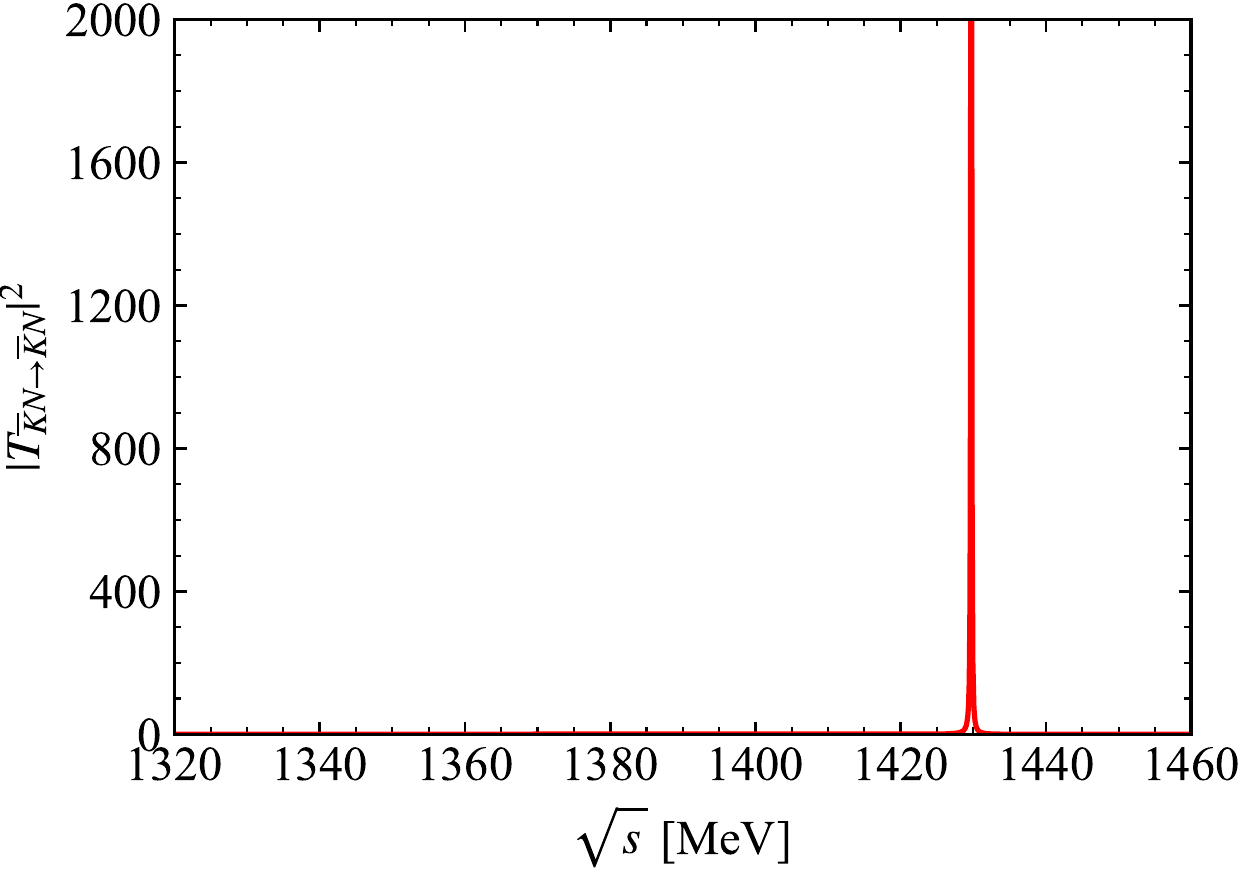} 
		\includegraphics[width=0.49\linewidth]{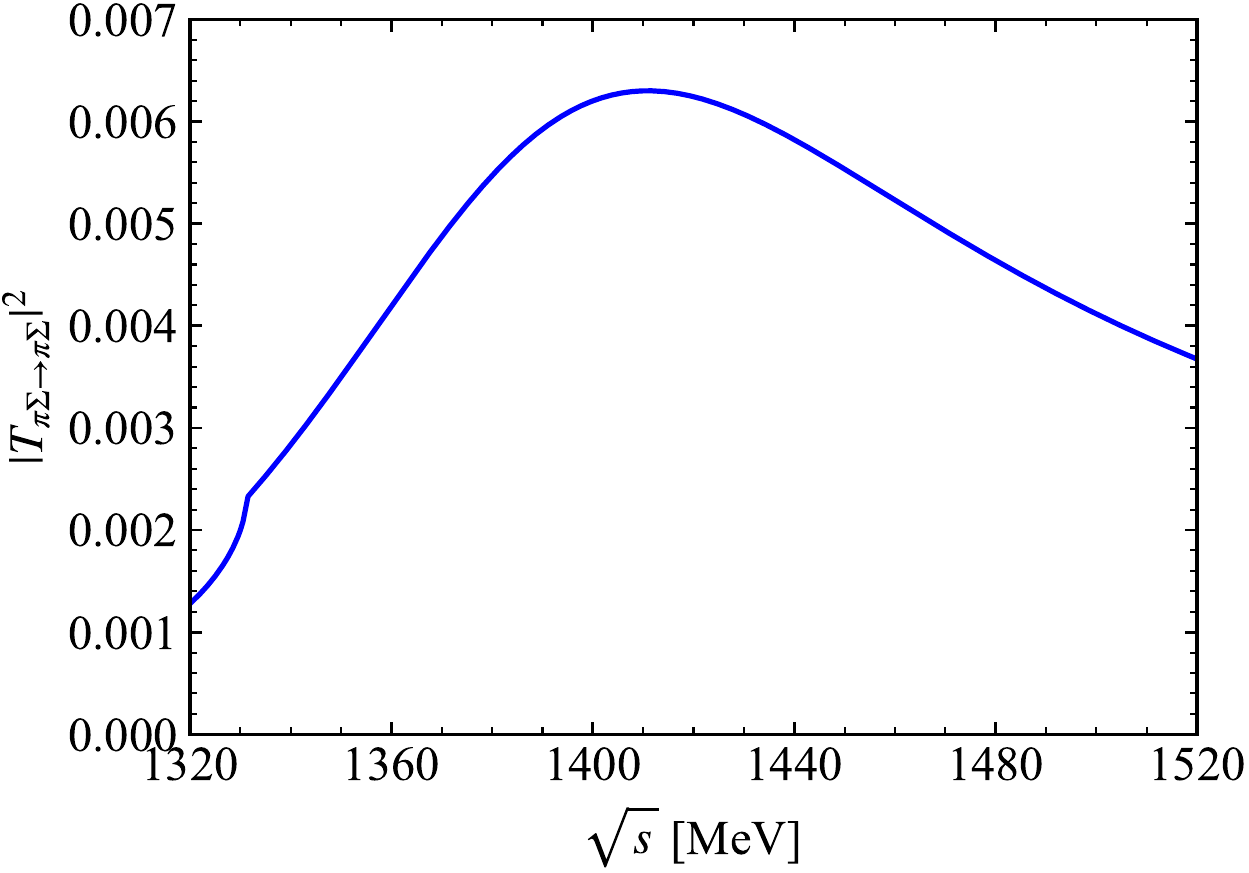} 
	\caption{Modulus squares of $T_{\bar{K}N\rightarrow\bar{K}N}$ (left) and $T_{\bar{K}N\rightarrow\pi\Sigma}$ (right) amplitudes in single channel interaction.}
	\label{fig:1T}
\end{figure}

For the sake of a comprehensive understanding of the pole structure of the $\Lambda(1405)$, we studied the single channel interaction, where the interactions for the channels $\bar{K} N$ and $\pi \Sigma$ were done individually without the coupled channel effect. The results of modulus squares for the scattering amplitudes are shown in Fig.~\ref{fig:1T}. From Fig.~\ref{fig:1T}, we can see a sharp peak with zero width in the $\bar{K}N$ channel and a wide bump structure in the $\pi\Sigma$ channel. Since we have found two poles which coupled to the $\bar{K} N$ and $\pi \Sigma$ channels in the coupled channel interactions, these structures were the corresponding ones to these channels in the singled channel interaction. In this case, there are only two Riemann sheets for each channel. As shown in Table~\ref{tab:1couplings}, we find one pole of $\sqrt{s_{p}}=1429.71$ located at the first Riemann sheet and below the threshold in the case of $\bar{K} N$ interaction, which is a bound state of $\bar{K} N$ with zero width for no channel to decay in the single channel interaction. This indicates that the interaction of $\bar{K} N$ is strong enough to bound a state. Whereas, the one of $\sqrt{s_{p}}=1380.75+80.95 i$ is appeared in the second Riemann sheet and above the threshold of $\pi \Sigma$ channel, which is a broad resonance for the weak potential of the $\pi \Sigma$ channel and analogous to the case of $f_0(500)$ in the $\pi\pi$ interaction, see the results of Ref.~\cite{Ahmed:2020kmp}. Hence, the pole of $\pi\Sigma$ has large decay width as in the case of the coupled channels, because it is above the $\pi\Sigma$ threshold and thus it still can decay to $\pi\Sigma$ channel. From these results, one can see clearly that two poles found in the coupled channel interactions above can be generated directly in the single channel interaction even without coupled channel effects, which is different from the case of $a_0(980)$ in the case of $\bar{K} K$ interactions in isospin $I=1$ sector as found in Ref.~\cite{Ahmed:2020kmp}.

Even though the coupling lose the physical meaning since there is no coupled channel interaction, we calculated the corresponding couplings, that are given in Table~\ref{tab:1couplings} for the purpose of evaluating the compositeness. The results of the compositeness are given in Table~\ref{tab:1compositeness}, which are consistent with the ones obtained in the case of coupled channel interactions above, see Table~\ref{tab:compositeness}. Note that the value of compositeness for the higher-mass pole is real for the real pole of a bound state.

\begin{table}
 \renewcommand{\arraystretch}{1.3}
	\centering
	\caption{Pole positions and couplings in single channel interaction.}	\label{tab:1couplings}
	\begin{tabular}{ccccc}
			\hline \hline 
			$\sqrt{s_{p}}$ & $g_{\bar{K} N}$ & $\left|g_{\bar{K}N}\right|$ & $g_{\pi\Sigma}$ & $\left|g_{\pi\Sigma}\right|$ \\
			\hline 
			$1380.75+80.95i$ & $-$ & $-$ & $1.97+1.45i$ & $2.44$ \\
			$1429.71$ & $1.81$ & $1.81$ & $-$ & $-$ \\
			\hline\hline    
	\end{tabular}
\end{table}

\begin{table}
 \renewcommand{\arraystretch}{1.3}
	\centering
	\caption{Compositeness of the poles in single channel interaction.}	\label{tab:1compositeness}
	\begin{tabular}{ccccc}
			\hline \hline 
			$\sqrt{s_{p}}$ & $(1-Z)_{\bar{K} N}$ & $\left|(1-Z)_{\bar{K}N}\right|$ & $(1-Z)_{\pi \Sigma}$ & $\left|(1-Z)_{\pi \Sigma}\right|$ \\
			\hline 
			$1380.75+80.95i$ & $-$ & $-$ & $0.30-0.35i$ & $0.45$ \\
			$1429.71$ & $0.89$ & $0.89$ & $-$ & $-$ \\
			\hline\hline    
	\end{tabular}
\end{table}

\begin{figure}
\begin{subfigure}{1\textwidth}  
		\centering
		\includegraphics[width=0.49\linewidth]{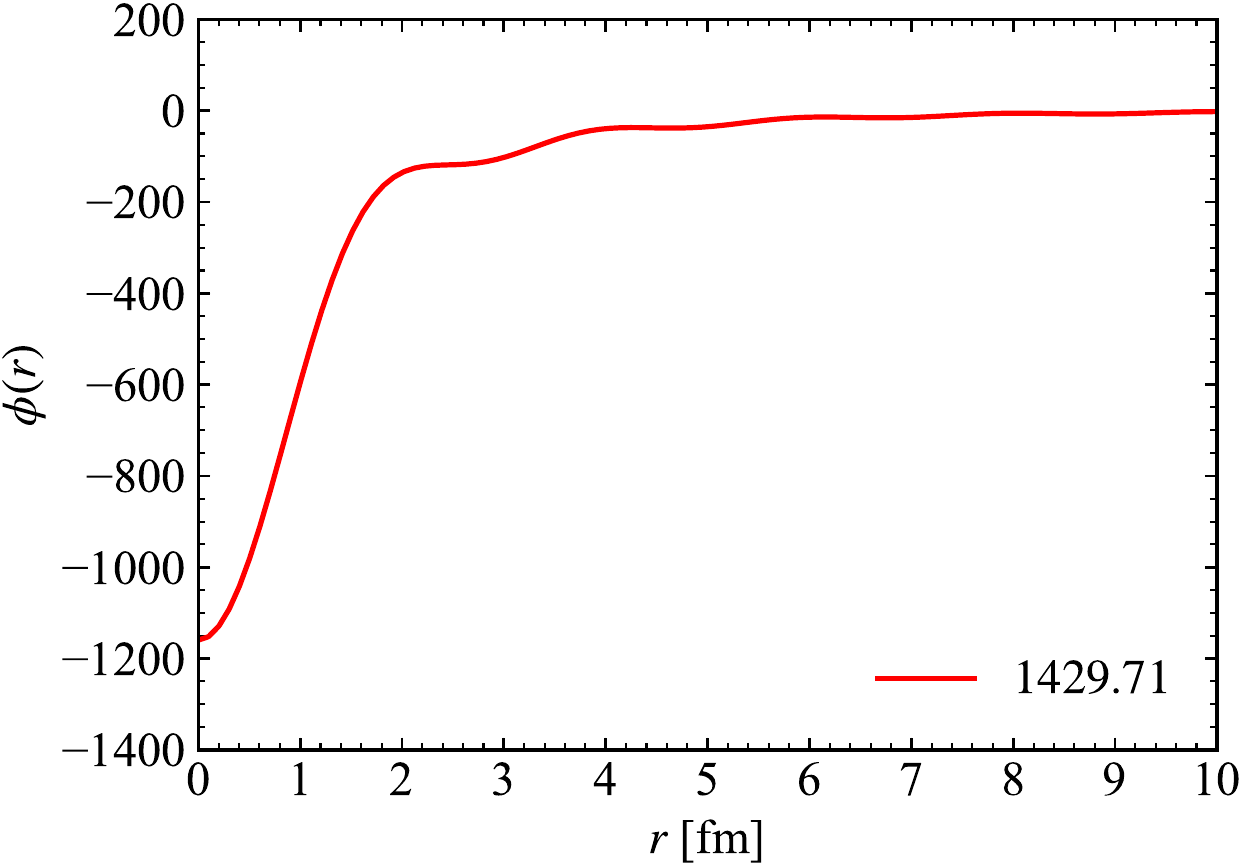} 
	\end{subfigure} 
	\begin{subfigure}{0.49\textwidth}  
	\centering
	\includegraphics[width=1\linewidth]{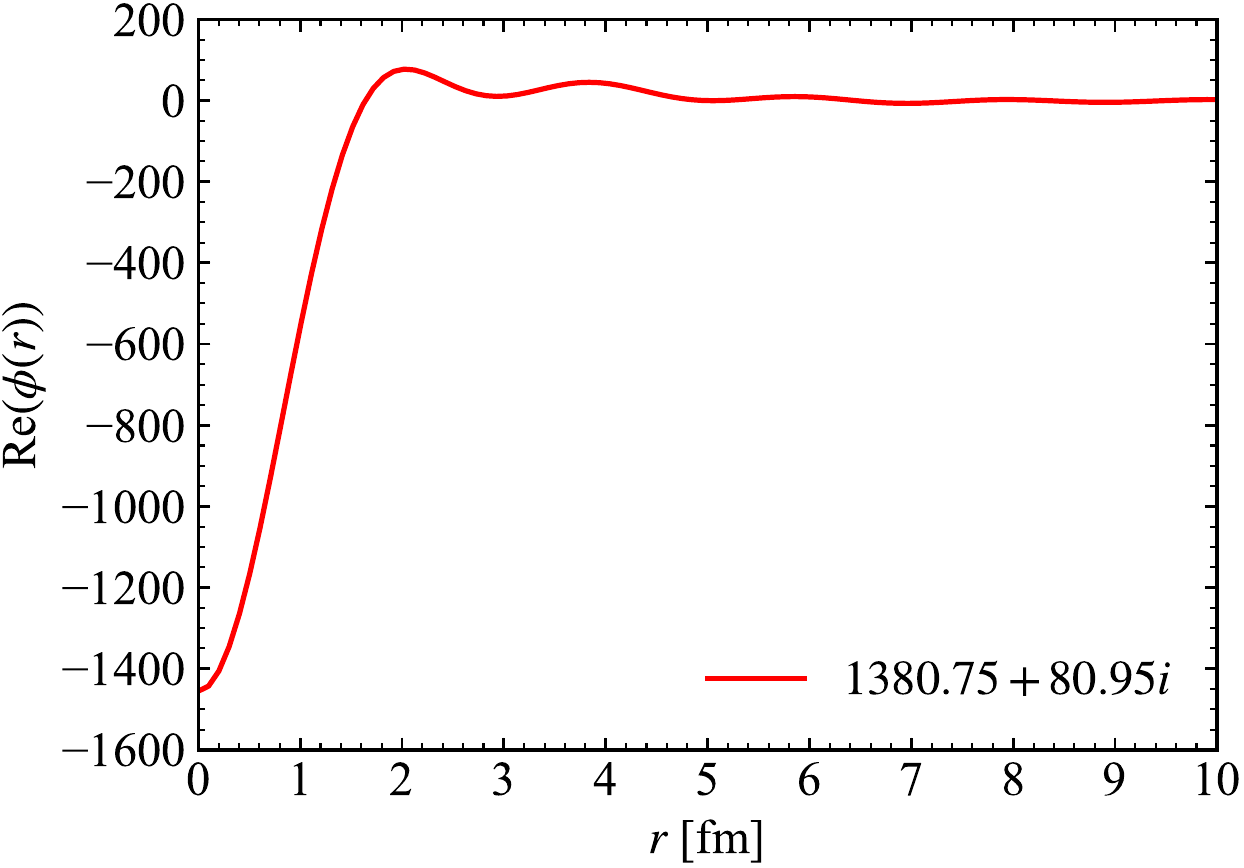} 
    \end{subfigure} 
	\begin{subfigure}{0.49\textwidth}  
	\centering
	\includegraphics[width=1\linewidth]{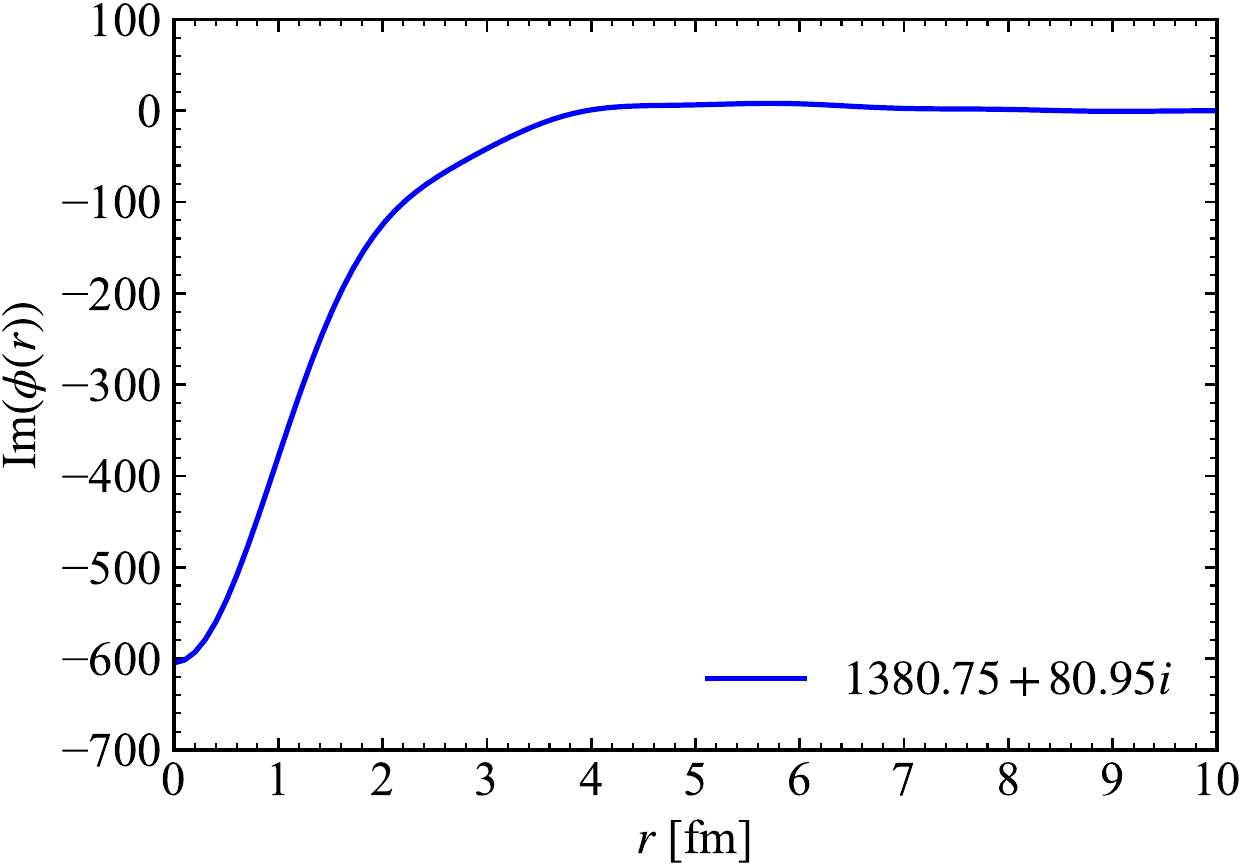} 
	\end{subfigure}
	\caption{Wave functions of the poles of $\bar{K}N$ (upper) and $\pi\Sigma$ (lower) in single channel approach.}
	\label{fig:1WF}
\end{figure}

To check whether the spatial distributions of the poles found in singled channel affected by the coupled channel effect or not, we also evaluated the corresponding wave functions, which are shown in Fig.~\ref{fig:1WF}. Now, due to the higher-mass pole is a bound state of no width, it has a real wave function. But, the behaviour of the function is still the same as the one in the coupled channel case. With the form factors obtained from the wave functions, the radii evaluated with Eq.~\eqref{eq:radii} for two poles are given in Table~\ref{tab:1radii}. Analogously, the radius for the higher-mass pole of $\bar{K}N$ is real now, where its value is $2.79$ fm and much greater than the ones $1.57$ fm or $1.58$ fm obtained in the case of coupled channel interactions, see Table~\ref{tab:radii}. But, the one for the lower-mass pole of $\pi \Sigma$ is $0.78$ fm, which is just slightly greater than the ones $0.67$ fm or $0.66$ fm in Table~\ref{tab:radii} and still kept in the the typical hadronic scale.

\begin{table}
 \renewcommand{\arraystretch}{1.3}
	\centering
	\caption{Radii of the poles evaluated with Eq.~\eqref{eq:radii} in single channel approach.}	\label{tab:1radii}
	\begin{tabular}{ccccc}
			\hline \hline 
			$\sqrt{s_{p}}$ & $\sqrt{\left\langle r^{2}\right\rangle_{\bar{K}N}}$ & $\mid \sqrt{\left\langle r^{2}\right\rangle_{\bar{K}N} }\mid$ & $\sqrt{\left\langle r^{2}\right\rangle_{\pi \Sigma}}$ & $\mid \sqrt{\left\langle r^{2}\right\rangle_{\pi \Sigma}}\mid$  \\
			\hline 
			$1380.75+80.95i$ & $-$ & $-$ & $(0.78+0.08i)$ fm&$0.78$ fm \\
			$1429.71$ & $2.79$ fm&$2.79$ fm&$-$ & $-$ \\
			\hline\hline    
	\end{tabular}
\end{table}

\section{Conclusions}

In the present work, to examine the properties of the $\Lambda(1405)$ state, we revisit the meson-baryon interactions with the chiral unitary approach, where we investigate the interactions of coupled channels (four coupled channels and two coupled channels) and singled channel (two singled channel) in isospin $I = 0$ and $S = -1$ sector. Both in the cases of four coupled channels and two coupled channels, two poles corresponding to the $\Lambda(1405)$ state are found in the second Riemann sheet. We determine the free parameters of $a_{\mu}$ in the loop functions with the constraint of some experimental data. With the Breit-Wigner parameterization, we show that how the $\Lambda(1405)$ peak found in the experimental invariant mass distribution of $\pi \Sigma$ can be demonstrated through the interference effects of two poles. In the case of the singled channel calculation, we find a pole from each channel interaction of $\bar{K} N$ and $\pi \Sigma$, which indicates that two poles can be dynamically generated from the single channel interactions even without the coupled channel effects. Based on these results, one can see that the nature of two poles is different.

Furthermore, we studied the couplings, the compositeness, the wave functions, and the mean-squared distance of these dynamically generated poles both in coupled channels and singled channel interactions. From the results of the couplings and compositeness, we find that the higher-mass pole is strongly coupled to $\bar{K} N$ channel and contributed mostly by the $\bar{K} N$ components, which is nearly up to 100\%. However, the lower-mass pole is tightly coupled to the $\pi \Sigma$ channel with some contributions from the $\bar{K} N$ interaction, where about 60\% of the pole component is $\pi \Sigma$ and about 20\% is $\bar{K} N$. With the wave functions obtained, we calculate the radii of two poles. From our results, the radius of the higher-mass pole is about 1.57 fm for the case of coupled channel interactions and 2.79 fm in the single channel interaction, which are larger than the typical hadronic scale. Whereas, the one for the lower-mass pole is about 0.66 fm and 0.78 fm for the coupled channel and single channel interactions, respectively, which are within the typical hadronic scale. Thus, the properties of the components and the sizes for the two poles are unique.

Eventually, based on these results obtained, one can conclude that the nature of two poles corresponding to the $\Lambda(1405)$ state is quite different. The higher-mass pole is a pure $\bar{K} N$ molecule, and the lower-mass one is a compositeness of main components of $\pi \Sigma$ and small part of $\bar{K} N$. Therefore, we achieve that the $\Lambda(1405)$ state is the superposition of two states with the same quantum number rather than only one ordinary particle.

\section*{Acknowledgments}

We thank Profs. Christoph Hanhart and Eulogio Oset for useful discussions, and acknowledge Prof. Eulogio Oset for careful reading the manuscript and valuable comments.

%%%%

\end{document}